\documentclass[apj,onecolumn]{emulateapj}

\usepackage{amsmath}
\usepackage{empheq}

\shortauthors{PANTONI ET AL.}
\shorttitle{NEW ANALYTIC SOLUTIONS FOR GALAXY EVOLUTION}
\slugcomment{ACCEPTED BY ApJ}

\begin{document}

\title{New Analytic Solutions for Galaxy Evolution:\\  Gas, Stars, Metals and Dust in local Early-Type Galaxies \\ and in their high-$z$ Starforming Progenitors}
\author{L. Pantoni\altaffilmark{1}, A. Lapi\altaffilmark{1,2,3,4}, M. Massardi\altaffilmark{5}, S. Goswami\altaffilmark{1}, L. Danese\altaffilmark{1,2}}
\altaffiltext{1}{SISSA, Via Bonomea 265, 34136 Trieste, Italy}\altaffiltext{2}{IFPU - Institute for fundamental physics of the Universe, Via Beirut 2, 34014 Trieste, Italy}\altaffiltext{3}{INFN-Sezione di Trieste, via Valerio 2, 34127 Trieste,  Italy}\altaffiltext{4}{INAF-Osservatorio Astronomico di Trieste, via Tiepolo 11, 34131 Trieste, Italy}\altaffiltext{5}{INAF, Istituto di Radioastronomia - Italian ARC, Via Piero Gobetti 101, I-40129 Bologna, Italy}

\begin{abstract}
We present a set of new analytic solutions aimed at self-consistently describing the spatially-averaged time evolution of the gas, stellar, metal, and dust content in an individual starforming galaxy hosted within a dark halo of given mass and formation redshift. Then, as an application, we show that our solutions, when coupled to specific prescriptions for parameter setting (inspired by in-situ galaxy-black hole coevolution scenarios) and merger rates (based on numerical simulations), can be exploited to reproduce the main statistical relationships followed by early-type galaxies and by their high-redshift starforming progenitors. Our analytic solutions allow to easily disentangle the diverse role of the main physical processes regulating galaxy formation, to quickly explore the related parameter space, and to make transparent predictions on spatially-averaged quantities. As such, our analytic solutions may provide a basis for improving the (subgrid) physical recipes presently implemented in theoretical approaches and numerical simulations, and can offer a benchmark for interpreting and forecasting current and future broadband observations of high-redshift starforming galaxies.
\end{abstract}

\section{Introduction}

Understanding the formation and evolution of galaxies in a cosmological context is one of the main challenge of modern astrophysics (e.g., Mo et al. 2010; Silk \& Mamon 2012; Maiolino \& Mannucci 2019). The issue is intrinsically very complex, since it involves many physical processes occurring on vastly different spatial, time, and energy scales.

The ultimate approach to address the problem in fine details would require the exploitation of intensive hydro-dynamical simulations (e.g., Dubois et al. 2014, 2016; Genel et al. 2014; Hopkins et al. 2014, 2018; Vogelsberger et al. 2014; Bekki 2013, 2015; Schaye et al. 2015; McAlpine et al. 2016, 2019; Zhukovska et al. 2016; Aoyama et al. 2017, 2018; McKinnon et al. 2017, 2018; Pallottini et al. 2017; Springel et al. 2018; Ricarte et al. 2019; Torrey et al. 2019; for a review and further references, see Naab \& Ostriker 2017). However, despite the considerable increase in resolution achieved recently, many of the physical processes relevant to the description of galaxy evolution constitute sub-grid physics and must be modeled via parametric recipes; in addition, a detailed exploration of the related parameter space or of different modeling prescriptions is often limited by exceedingly long computational times.

An alternative route to investigate the issue relies on semi-analytic models (e.g., Lacey \& Cole 1993; Kauffmann et al. 1993; Cole et al. 2000; De Lucia et al. 2014, 2017; Croton et al. 2006; Somerville et al. 2008; Arrigoni et al. 2010; Benson 2012;  Porter et al. 2014; Cousin et al. 2016; Hirschmann et al. 2016; Lacey et al. 2016; Fontanot et al. 2017; Popping et al. 2017; Collacchioni et al. 2018; Lagos et al. 2018; Forbes et al. 2019; for a review and further references, see Somerville \& Dave 2015). These are based on dark matter (DM) merger trees extracted from, or gauged on, $N-$body simulations, while the physics relevant for galaxy evolution inside dark halos is modeled via several parametric expressions partly set on a number of (mainly local) observables. Such models are less computationally expensive than hydrodynamic simulations and allow to disentangle more clearly the relative role of the various physical processes at work; however, the considerable number of fudge parameters can lead to degenerate solutions and limit somewhat the predictive power, especially toward high redshift.

Finally, some specific issues related to the global evolution of the baryonic content in galaxies can be tackled with analytic models, i.e., models with analytic solutions (e.g., Schmidt 1963; Talbot \& Arnett 1971; Tinsley 1974; Pagel \& Patchett 1975; Hartwick 1976; Chiosi 1980; Matteucci \& Greggio 1986; Edmunds 1990; Dwek 1998; Hirashita 2000, Hirashita et al. 2015; Chiappini et al. 2001; Draine 2003, 2011; Inoue 2003; Greggio 2005; Naab \& Ostriker 2006; Erb 2008; Zhukovska et al. 2008; Valiante et al. 2009; Bouch\'e et al. 2010; Dwek \& Cherchneff 2011; Dav\'e et al. 2012; Asano et al. 2013; Lilly et al. 2013; De Benassutti et al. 2014; Dekel \& Mandelker 2014; Forbes et al. 2014a; Pipino et al. 2014; Feldmann 2015; Mancini et al. 2015; Moll\'a et al. 2015; Recchi \& Kroupa 2015; Rodriguez-Puebla et al. 2016; Andrews et al. 2017; Gioannini et al. 2017; Spitoni et al. 2017; Weinberg et al. 2017; Vincenzo et al. 2017; Grisoni et al. 2018; Imara et al. 2018; Tacchella et al. 2018a; Dekel et al. 2019; for a review and further references, see Matteucci 2012). These are necessarily based on approximate and spatially/time-averaged descriptions of the most relevant astrophysical processes; however, the transparent, handy and predictive character of analytic solutions often pay off.

In the present paper we develop a set of new analytic solutions aimed at describing the spatially-averaged time evolution of the gas, stellar, metal, and dust content in an individual starforming galaxy hosted within a dark halo of given mass and formation redshift.
Our basic framework pictures a galaxy as an open, one-zone system comprising three interlinked
mass components: a reservoir of warm gas subject to cooling and condensation toward the central regions; cold gas fed by infall and depleted by star formation and stellar feedback (type-II supernovae [SNe] and stellar winds); stellar mass, partially restituted to the cold phase by stars during their evolution. Remarkably, the corresponding analytic solutions for the metal enrichment history of the cold gas and stellar mass are self-consistently derived using as input the solutions for the evolution of the mass components; the metal equations includes effects of feedback, astration, instantaneous production during star formation, and delayed production by type-I$a$ SNe, possibly following a specified delay time  distribution.  Finally,  the  dust  mass  evolution  takes  into  account  the  formation  of grain  cores  associated  to  star  formation,  and  of  the  grain  mantles  due  to  accretion  onto pre-existing cores; astration of dust by star formation and stellar feedback, and spallation by SN shockwaves are also included.

We then apply our analytic solutions to describe the formation of spheroids/early-type
galaxies (ETGs) and the evolution of their starforming progenitors.
To this purpose, we couple our solutions to two additional ingredients:
(i) specific prescriptions for parameter setting, inspired by in-situ galaxy-black hole coevolution scenarios of ETG formation; (ii) estimates of the halo and stellar mass growth by mergers, computed on the basis of the
merger rates from state-of-the-art numerical simulations.
We then derive and confront with available data a bunch of fundamental spatially-averaged quantities: the star formation efficiency, the SFR, the gas mass, the gas metallicity, the stellar metallicity, the [$\alpha$/Fe] ratio, the dust mass, and the outflowing gas metallicity as a function of the observed stellar mass.

The paper is organized as follows: in Sect. \ref{sec|An_sol} we present new analytic solutions for the time evolution of the gas and stellar masses (Sect. \ref{sec|GasStars}), metals (Sect. \ref{sec|Met}), dust (Sect. \ref{sec|Dust}), and outflowing matter (Sect. \ref{sec|FedOut}); our solutions are compared with classic analytic models in Appendix A. In Sect. \ref{sec|Application} we apply our solutions to high-redshift starforming galaxies: we first provide physical prescriptions to set the parameters ruling infall and star formation (Sect. \ref{sec|biascoll}), feedback (Sect. \ref{sec|feedbacks}), metal and dust production/restitution (Sect. \ref{sec|otherparams}); we then discuss how to include the halo and stellar mass growth after formation (Sect. \ref{sec|merging_sect}) and how to compute observable quantities at a given observation redshift by averaging over different formation redshifts (Sect. \ref{sec|average_sect}). In Sect. \ref{sec|Results} we present a comparison between our results and available observations concerning the evolution of individual galaxies (Sect. \ref{sec|timevo}), the star formation efficiency (Sect. \ref{sec|fstar_sect}), the galaxy main sequence (Sect. \ref{sec|MS_sect}), the gas mass (Sect. \ref{sec|gasmass_sect}), the dust mass (Sect. \ref{sec|dustmass}), the gas metallicity (Sect. \ref{sec|gasmetal}), the stellar metallicity and $\alpha-$enhancement (Sect. \ref{sec|starmetallicity_sect}), and the outflowing gas metallicity (Sect. \ref{sec|outmetals}). In Section \ref{sec|summary} we summarize our approach and main findings.

Throughout this work, we adopt the standard flat $\Lambda$CDM cosmology
(Planck Collaboration 2018) with rounded parameter values: matter density $\Omega_M = 0.32$, dark energy density $\Omega_\Lambda=0.63$, baryon density $\Omega_b = 0.05$, Hubble constant $H_0 = 100\,h$ km s$^{-1}$ Mpc$^{-1}$  with $h = 0.67$, and mass variance $\sigma_8 = 0.81$ on a scale of $8\, h^{-1}$ Mpc. In addition, we use the widely adopted Chabrier (2003, 2005) initial mass function (IMF) with shape $\phi(\log m_\star)\propto \exp[-(\log m_\star-\log 0.2)^2/2\times 0.55^2]$ for $m_\star\la 1\, M_\odot$ and $\phi(\log m_\star)\propto m_\star^{-1.35}$ for $m_\star\ga 1\, M_\odot$, continuously joint at $1\, M_\odot$ and normalized as $\int_{0.1\, M_\odot}^{100\, M_\odot}{\rm d}m_\star\, m_\star\, \phi(m_\star)=1\, M_\odot$. Finally, a value $Z_\odot\approx 0.014$ for the solar metallicity is adopted, corresponding to $12+\log[O/H]_{\odot}=8.69$ (see Allende Prieto et al. 2001).

\section{Analytic solutions for individual galaxies}\label{sec|An_sol}

In this section we present new analytic solutions for the time evolution of the mass, metal and dust components in high-$z$ starforming galaxies, and in particular in the progenitors of ETGs.
These are derived from a quite general framework, meant to capture the main physical processes regulating galaxy formation on a spatially-averaged ground. The most relevant solutions are highlighted with a box; a comparison of these with classic analytic models, like the closed/leaky-box and gas regulator, is provided in Appendix A.

\subsection{Gas and stars}\label{sec|GasStars}

We consider a one-zone description of a galaxy with three interlinked mass components: the infalling gas mass $M_{\rm inf}$, the cold gas mass $M_{\rm cold}$, and the stellar mass $M_{\star}$. The evolution of these components is described by the following system of ordinary differential equations:
\begin{equation}\label{eq|basics}
\left\{
\begin{aligned}
\dot M_{\rm inf} &= -\frac{M_{\rm inf}}{\tau_{\rm cond}}~,\\
\\
\dot M_\star &=\frac{M_{\rm cold}}{\tau_\star}~,\\
\\
\dot M _{\rm cold} &= \dot M_{\rm inf} - \gamma\, \dot M_\star~,
\end{aligned}
\right.
\end{equation}
where $\gamma\equiv 1-\mathcal{R}+\epsilon_{\rm out}$. These equations prescribe that the infalling gas mass $M_{\rm inf}$ cools and condenses into the cold gas phase $M_{\rm cold}$ over a characteristic timescale $\tau_{\rm cond}$; then the stellar mass $M_{\rm \star}$ is formed from the cold mass $M_{\rm cold}$ at a rate $\dot M_{\rm \star}$ over a characteristic timescale $\tau_{\star}=\tau_{\rm cond}/s$; the cold gas mass is further replenished at a rate $\mathcal{R}\,\dot M_{\rm \star}$ by stellar recycling, where $\mathcal{R}$ is the return fraction of gaseous material from stellar evolution, and it is removed at a rate $\epsilon_{\rm out}\, \dot M_{\rm \star}$ by outflows driven from type-II SN explosions and stellar winds, where $\epsilon_{\rm out}$ is the mass loading factor. Note that in the above Eqs.~(\ref{eq|basics}) the quantity $M_\star(\tau)\equiv \int_0^\tau{\rm d}\tau' \dot M_\star(\tau')$ represents the integral of the SFR over the galactic age, while the true relic stellar mass after the loss due to stellar evolution is $M_\star^{\rm relic}\simeq (1-\mathcal{R})\, M_\star$. We adopt an IMF $\phi(m_\star)$ uniform in space and constant in time, and assume the instantaneous mixing (gas is well mixed at anytime) and instantaneous recycling (stars with mass $m_\star\ga 1\, M_\odot$ die as soon as they form, while those with $m_\star\la 1\, M_\odot$ live forever) approximations, so that the recycled fraction (fraction of a stellar population not locked into long-living dark remnants) can be computed as
\begin{equation}
\mathcal{R}\equiv \int_{1\, M_\odot}^{100\, M_\odot}{\rm d}m_\star\, (m_\star-m_{\rm rem})\, \phi(m_\star)
\end{equation}
where $m_{\rm rem}(m_\star)$ is the mass of the remnants; for our fiducial Chabrier (2003, 2005) IMF, the recycling fraction amounts to $\mathcal{R}\approx 0.45$. Standard initial conditions for the above system of equations read $M_{\mathrm{inf}}(0)=f_{\rm inf}\, M_{\rm b}$ and $M_{\mathrm{cold}}(0)=M_{\mathrm{\star}}(0)=0$; here $M_b=f_b\, M_{\rm H}$ is the baryonic mass originally present in the host halo with mass $M_{\rm H}$, while $f_{\rm inf}=M_{\rm inf}/f_b\,M_{\rm H}$ is the fraction of such a mass that can effectively cool and inflow toward the inner regions of the halo over the timescale $\tau_{\rm cond}$. We will discuss prescriptions for setting these parameters, apt for high-$z$ starforming galaxies and in particular for ETG progenitors, in Sect.~\ref{sec|Application}; for the reader convenience, we anticipate here that typical values for a halo with mass $M_{\rm H}\approx 10^{12}\, M_\odot$ formed at $z_{\rm form}\approx 2$ turn out to be $\tau_{\rm cond}\approx$ a few $10^8$ yr, $s\approx 3$, $f_{\rm inf}\la 1$ and $\epsilon_{\rm out}\la 2$.

A few caveats on the general structure of Eqs.~(\ref{eq|basics}) are in order. First, being our main focus to derive analytic solutions for high-redshift starforming galaxies, and in particular for the progenitors of local spheroids/ETGs, we have not included in the equations above a number of processes that can be otherwise relevant for the evolution of local disk-dominated/spiral galaxies, such as galactic fountains and radial gas flows (e.g., Bregman 1980; Lacey \& Fall 1985; Pitts \& Tayler 1989; Spitoni et al. 2008, 2013; Fu et al. 2013; Forbes et al. 2014b; Pezzulli \& Fraternali 2016; Stevens et al. 2016, 2018; Stevens \& Brown 2017), differential galactic winds (e.g., Pilyugin 1993; Marconi et al. 1994; Recchi et al. 2008), multi-zonal structures and stellar mixing (see Kubrik et al. 2015; Spitoni et al. 2015; Grisoni et al. 2018).

Second, as it will be discussed in Sect.~\ref{sec|feedbacks}, the star formation timescale and the duration of the main star formation episode in high $z\ga 1$ starforming galaxies, and in particular in ETG progenitors, are typically of order from a few $10^8$ to $10^9$ yr (e.g., Thomas et al. 2005; Gallazzi et al. 2006, 2014); moreover, most of the star formation process occurs in a central compact region of size around a few kpcs (e.g., Scoville et al. 2014, 2016; Ikarashi et al. 2015; Simpson et al. 2015; Straatman et al. 2015; Spilker et al. 2016; Tadaki et al. 2017a,b; Lang et al. 2019). Given that, one can safely neglect in Eqs.~(\ref{eq|basics}) additional terms describing the growth of the host DM halo and of the stellar content due to accretion and/or mergers, that typically occur over cosmological timescales of several Gyrs and large spatial scales of order from several tens kpcs (stellar mass addition by galaxy mergers) to several hundreds kpcs (growth of DM halo and gaseous baryon reservoir); nonetheless, these mass additions will become relevant in the long-term evolution of ETG progenitors toward the present and will be included in our computations, as detailed in Sect.~\ref{sec|merging_sect}.

Third, note that a classic way to link the SFR and the cold gas mass, adopted by many analytic models focused on disk galaxies, is the classic Schmidt-Kennicutt (Schmidt 1959; Kennicutt 1998) law; this prescribes $\dot \Sigma_\star\propto \Sigma_{\rm cold}^{1.4}$ in terms of the stellar and gas disk surface densities $\Sigma_\star$ and $\Sigma_{\rm cold}$. However, in the equations above we have adopted instead a spatially-averaged star formation law $\dot M_\star\propto M_{\rm cold}$ linking linearly the SFR and the total cold gas mass; this is indicated by recent observations of local starbursts and high-redshift $z\ga 1$ starforming galaxies (see Krumholz et al. 2012; Scoville et al. 2016, 2017), and is also suggested for local disk galaxies by spatially resolved observations of dense gas in molecular clouds (e.g., Bigiel et al. 2008; Lada et al. 2010). We further note that the star formation law $\dot M_\star = M_{\rm cold}/\tau_\star$ has been written in terms of the total cold gas mass, but it can be equivalently expressed as $\dot M_\star = f_{\rm H2}\,M_{\rm cold}/\tau_{\star,\rm H2}$ in terms of the molecular gas fraction $f_{\rm H2}$ by simply redefining the star formation timescale $\tau_{\star}=\tau_{\star,\rm H2}/f_{\rm H2}$ (see Feldmann 2015).

The above system of linear equations can be easily solved analytically in the form
\begin{empheq}[box=\fbox]{align}\label{eq|basicsol}
\nonumber\\
\left\{
\begin{aligned}
M_{\mathrm{inf}}(x) &= f_{\rm inf}\, M_{\mathrm{b}}\, e^{-x}~,\\
\\
M_{\mathrm{cold}}(x) &= \frac{f_{\rm inf}\, M_{\mathrm{b}}}{s\,\gamma-1}\, \left[e^{-x}-e^{-s\, \gamma\, x}\right]~,\\
\\
M_{\mathrm{\star}}(x) &= \frac{f_{\rm inf}\, M_{\mathrm{b}}\, s}{s\,\gamma-1}\, \left[1-e^{-x}-\frac{1}{s\,\gamma}\,\left(1-e^{-s\, \gamma\, x}\right)\right]~,
\end{aligned}
\right.\\
\nonumber
\end{empheq}
where $x\equiv \tau/\tau_{\rm cond}$ is a dimensionless time variable normalized to the condensation timescale, and $s\equiv \tau_{\rm cond}/\tau_{\rm \star}$ is the ratio between the condensation and the star formation timescales. Note that the above solution is physically meaningful (specifically, the cold and stellar masses are non-negative for any $x$) whenever $s\gamma > 1$. Since $1/\gamma=1/(1-\mathcal{R}+\epsilon_{\rm out})\la 1/(1-\mathcal{R})\la 2$ it is sufficient that $s\ga 2$; as it will be discussed in Sects.~\ref{sec|biascoll}-\ref{sec|feedbacks} and shown in Fig.~\ref{fig|params}, for ETG progenitors typical values $s\ga 3$ apply at any mass and redshift, so that the above condition is regularly met.
In a nutshell, according to Eqs.~(\ref{eq|basicsol}) the infalling gas mass declines exponentially with time, while the cold gass mass (hence the SFR) features an initial growth, then attains a maximum and eventually declines exponentially. Our analytic solution for the gas mass is compared with that from classic analytic models in Appendix A.

It is instructive to examine the initial behavior of the solutions for $\tau\ll \tau_{\rm cond}$, that reads
\begin{equation}
\left\{
\begin{aligned}
M_{\rm inf} &\simeq f_{\rm inf}\, M_{\mathrm{b}}\, \left(1- \frac{\tau}{\tau_{\rm cond}}\right)~,\\
\\
M_{\rm cold} &\simeq f_{\rm inf}\, M_{\mathrm{b}}\, \left(\frac{\tau}{\tau_{\rm cond}}\right)~,\\
\\
M_{\rm \star} &\simeq \frac{s\, f_{\rm inf}\, M_{\mathrm{b}}}{2}\,\left(\frac{\tau}{\tau_{\rm cond}}\right)^2~,
\end{aligned}
\right.
\end{equation}
where the dimensionless independent variable $x$ has been re-expressed as $\tau/\tau_{\rm cond}$. In addition, the maximum of cold gas mass (and SFR) occurs for
\begin{equation}
\tau_{\rm max} = \tau_{\rm cond}\,\ln\left[(s\,\gamma)^{1/(s\,\gamma-1)}\right]~
\end{equation}
when the mass of the various gas components writes
\begin{equation}
\left\{
\begin{aligned}
M_{\rm inf}(\tau_{\rm max}) &\simeq f_{\rm inf}\, M_{\mathrm{b}}\, (s\,\gamma)^{-1/(s\, \gamma-1)}~,\\
\\
M_{\rm cold}(\tau_{\rm max}) &\simeq f_{\rm inf}\, M_{\mathrm{b}}\, (s\,\gamma)^{-s\, \gamma/(s\, \gamma-1)}~,\\
 \\
M_{\rm \star}(\tau_{\rm max}) &\simeq \frac{f_{\rm inf}\, M_{\mathrm{b}}}{\gamma}\,\left[1-(1+s\, \gamma)\, (s\, \gamma)^{-s\, \gamma/(s\, \gamma-1)}\right]~.
\end{aligned}
\right.
\end{equation}
Finally, for $\tau\gg \tau_{\rm cond}$ the solutions behave as
\begin{equation}
\left\{
\begin{aligned}
M_{\mathrm{inf}} &\simeq f_{\rm inf}\, M_{\mathrm{b}}\, e^{-\tau/\tau_{\rm cond}}~,\\
\\
M_{\mathrm{cold}} &\simeq \frac{f_{\rm inf}\, M_{\mathrm{b}}}{s\,\gamma-1}\, e^{-\tau/\tau_{\rm cond}}~,\\
\\
M_{\mathrm{\star}} &\simeq \frac{f_{\rm inf}\, M_{\mathrm{b}}}{\gamma}~.
\end{aligned}
\right.
\end{equation}

These expressions highlight a few interesting facts. First, the overall time behavior of the cold gas mass and of the SFR $\dot M_\star(\tau)\propto M_{\rm cold}(\tau)$ is very similar to the empirical delayed exponential shape $\dot M_\star\propto\tau^\kappa\, e^{-\tau/\tau_{\rm cond}}$ with $\kappa\la 1$, which is routinely used to describe the star formation histories and to interpret the spectral energy distribution for high-$z$ starforming galaxies and for the progenitors of local spheroids (see Papovich et al. 2011; Moustakas et al. 2013; Steinhardt et al. 2014; da Cunha et al. 2015; Citro et al. 2016; Cassar\'a et al. 2016; Boquien et al. 2019); thus our solutions may provide a physical basis to that empirical shape. Second, after the peak of the SFR for $\tau\ga \tau_{\rm max}$ the infall rate and the SFR are proportional, such that $\dot M_{\rm inf}\propto -(s\gamma-1)\dot M_\star$; this explains why models where a similar proportionality is assumed a priori (e.g., Matteucci \& Chiosi 1983; Matteucci 2012) produce results not substantially different from those with a generic exponential infall (see Recchi et al. 2008). Third, the specific star formation rate sSFR $\equiv \dot M_\star/M_\star$ is a monotonic function of the galaxy age, since at early times it behaves like sSFR $\simeq 2/\tau$ and at late times as sSFR $\simeq s\gamma\, e^{-\tau/\tau_{\rm cond}}/(s\gamma-1)$; thus a selection in sSFR is equivalent to a selection in galaxy age. Lastly, the true relic stellar mass after the loss due to stellar evolution is
\begin{equation}
M_{\rm \star}^{\rm relic}\simeq (1-\mathcal{R})\,M_{\rm \star} = \frac{1-\mathcal{R}}{1-\mathcal{R}+\epsilon_{\rm out}}\, f_{\rm inf}\, M_{\mathrm{b}}~;
\end{equation}
thus, in absence of any outflows $\epsilon_{\rm out}\approx 0$, all the available (infalling) baryons would be converted into stars.

\newpage

\subsection{Metals}\label{sec|Met}

We now turn to discuss the time evolution of the metallicity in cold gas and in stellar mass, first focusing on instantaneous production of $\alpha-$elements by type-II SNe and stellar winds, and then turning to the delayed production of iron group elements by type-I$a$ SNe.

\subsubsection{Cold gas metallicity from instantaneously produced elements}

The time evolution of the metallicity $Z_{\rm cold}$ in cold gas mass contributed by instantaneously produced chemical elements is described as
\begin{equation}\label{eq|Zcold}
\frac{{\rm d}}{{\rm d}\tau}(M_{\rm cold}\, Z_{\rm cold})=-\gamma\,\dot{M}_{\rm \star}\, Z_{\rm cold}+{y_Z\,(1-\mathcal{R})}\, \dot{M}_{\rm \star}~.
\end{equation}
The equation above prescribes that the mass of metals in cold gas, $M_{\rm cold}\,Z_{\rm cold}$, evolves because of instantaneous metal production at a rate $y_Z\, (1-\mathcal{R})\, \dot M_{\rm \star}$, outflow depletion at a rate $\epsilon_{\rm out}\, \dot M_{\rm \star}\,Z_{\rm cold}$, and
astration (metal mass locking into stellar remnants) at a rate $(1-\mathcal{R})\,\dot M_{\rm \star}\, Z_{\rm cold}$; we have neglected the infalling gas metallicity, that is assumed to be primordial (but can be otherwise included very easily in the analytic solutions). Under the instantaneous mixing and recycling approximation, the metal production yield is given by
\begin{equation}
y_Z\equiv \frac{1}{1-\mathcal{R}}\,\int_{1\, M_\odot}^{100\, M_\odot}{\rm d}m_\star\, m_\star\, p_{Z,\star}\,\phi(m_\star)
\end{equation}
where $p_{Z,\star}$ is the mass-fraction of newly synthesized metals by the star of initial mass $m_\star$; with this definition relative to $1-\mathcal{R}$, the yield $y_Z$ represents the ratio between the mass of heavy elements ejected by a stellar generation and the mass locked up in remnants.

In many previous analytic models, to solve the chemical evolution equation an empirical shape of the SFR is adopted; remarkably, here we instead use the self-consistent solutions for the time evolution of the infalling and cold gas masses, recasting  Eq.~(\ref{eq|Zcold}) into the form
\begin{equation}
\dot{Z}_{\rm cold} = -\frac{Z_{\rm cold}}{\tau_{\rm cond}}\, \frac{s\, \gamma-1}{1-e^{-(s\, \gamma-1)\, \tau/\tau_{\rm cond}}}+\frac{y_Z\, (1-\mathcal{R})\, s}{\tau_{\rm cond}}~.
\end{equation}
with initial condition $Z_{\rm cold}(0)=0$. The corresponding analytic solution for the cold gas metallicity reads
\begin{empheq}[box=\fbox]{align}\label{eq|Zcold_inst}
\nonumber\\
Z_{\rm cold}(\tau) = \bar Z_{\rm cold}\, \left[1-\frac{(s\,\gamma-1)\, x}{e^{(s\, \gamma-1)\, x}-1}\right]~,\\
\nonumber
\end{empheq}
where
\begin{equation}
\bar Z_{\rm cold} = \frac{s\, y_Z\, (1-\mathcal{R})}{s\, \gamma-1}~
\end{equation}
represents the asymptotic value for $\tau\gg \tau_{\rm cond}$. It turns out that the evolution of the gas metallicity is very rapid, so that it attains values $\ga Z_{\odot}/10$ in a quite short timescale $\tau\la \tau_{\rm cond}/10\approx$ a few $10^7$ yr; this will be relevant for the metal and dust enrichment of primordial galaxies and quasars at high redshift (see Sect.~\ref{sec|gasmetal}.

It is instructive to look at the initial behavior of the gas metallicity for $\tau\ll \tau_{\rm cond}$, that reads
\begin{equation}
Z_{\rm cold}(\tau) \simeq \frac{s\,y_Z\, (1-\mathcal{R})}{2}\, \frac{\tau}{\tau_{\rm cond}}~;
\end{equation}
the resulting evolution is thus almost linear with galactic age until a saturation to the final, stationary value $\bar Z_{\rm cold}$ takes place. Interestingly, the cold gas metallicity evolution at early times can be expressed in terms of the sSFR $\equiv \dot M_\star/M_\star$, such that it behaves like $Z_{\rm cold}\simeq s\,y_Z\, (1-\mathcal{R})\,/$sSFR$ \tau_{\rm cond}\propto M_\star/\dot M_\star$; thus a selection based on small $M_\star$ or large sSFR will tend to pick up objects for which the gas metallicity scales inversely with the SFR. On the other hand, at late times the gas metallicity saturates to a constant value $\bar Z_{\rm cold}$; thus a selection based on large $M_\star$ or small sSFR will tend to pick up galaxies for which the cold gas metallicity is uncorrelated with the SFR. This is the essence of the fundamental metallicity relation, established observationally by Mannucci et al. (2010; see also Lara-Lopez et al. 2013).

Our analytic solution for the gas metallicity is compared with that from classic analytic models in Appendix A.

\subsubsection{Cold gas metallicity from delayed-produced elements}\label{sec|del_Z_gas_subsect}

We can also deal with a delayed production of metals, related to the iron-group elements, due to type-I$a$ SNe. To a good approximation one can decompose $Z_{\rm cold}\simeq Z^{\Delta=0}_{\rm cold}+Z_{\rm cold}^{\Delta}$ into an instantaneous production component $Z^{\Delta=0}_{\rm cold}$ as described in the previous Section, and a delayed production component $Z^{\Delta}_{\rm cold}$, governed by the equation
\begin{equation}
\frac{{\rm d}}{{\rm d}\tau}(M_{\rm cold}\, Z^{\Delta}_{\rm cold})=-\gamma\,\dot{M}_{\rm \star}\, Z_{\rm cold}^{\Delta}+{y_Z^{\Delta}\,(1-\mathcal{R})}\, \dot{M}_{\rm \star}(\tau)\, \frac{\dot{M}_{\rm \star}(\tau-\Delta)}{\dot{M}_{\rm \star}(\tau)}~,
\end{equation}
with $y_Z^{\Delta}$ the stellar yield for delayed metals, and $\Delta$ the delay time for the enrichment by type-I$a$ SNe, that for the moment we consider fixed to some particular value. Notice that this equation is very similar to that for instantaneously produced metals, apart from the factor $\dot{M}_{\rm \star}(\tau-\Delta)/ \dot{M}_{\rm \star}(\tau)$ in the last term on the r.h.s., which accounts for the delayed contribution of type-I$a$ SNe in polluting the cold medium.

Using the self-consistent solutions for the SFR, the evolution equation for the delayed metallicity $Z_{\rm cold}^\Delta$ is found:
\begin{equation}
\dot{Z}_{\rm cold}^{\Delta} = -\frac{Z_{\rm cold}^{\Delta}}{\tau_{\rm cond}}\, \frac{s\, \gamma-1}{1-e^{-(s\, \gamma-1)\, \tau/\tau_{\rm cond}}}+\frac{y_Z^{\Delta}\, (1-\mathcal{R})\, s}{\tau_{\rm cond}}\, \frac{e^{\Delta/\tau_{\rm cond}}-e^{s\,\gamma\, \Delta/\tau_{\rm cond}}\, e^{-(s\, \gamma-1)\, \tau/\tau_{\rm cond}}}{1-e^{-(s\, \gamma-1)\, \tau/\tau_{\rm cond}}}~.
\end{equation}
with initial condition $Z_{\rm cold}^\Delta(\tau)=0$ for $\tau<\Delta$. The corresponding analytic solution reads
\begin{empheq}[box=\fbox]{align}\label{eq|Zcold_del}
\nonumber\\
Z_{\rm cold}^\Delta(\tau) = \left\{
\begin{aligned}
\bar Z_{\rm cold}^\Delta &\, \left[1-\frac{e^{(s\,\gamma-1)\, x_\Delta}\,[1+(s\,\gamma-1)\, (x-x_\Delta)]-1}{e^{(s\, \gamma-1)\, x}-1}\right] & {\rm for}~~~~x\geq x_{\Delta}~,\\
\\
& 0 &  {\rm for}~~~~x<x_{\Delta}
\end{aligned}
\right.\\
\nonumber
\end{empheq}
where $x_\Delta\equiv \Delta/\tau_{\rm cond}$ and $\bar Z_{\rm cold}^\Delta$ is the delayed metallicity asymptotic behavior for $\tau\gg \tau_{\rm cond}$:
\begin{equation}
\bar Z_{\rm cold}^\Delta = \frac{s\, y_Z^\Delta\, (1-\mathcal{R})\, \,e^{\Delta/\tau_{\rm cond}}}{s\, \gamma-1}~.
\end{equation}
Note that, as it should be, for $\Delta=0$ the time dependence in Eq.~(\ref{eq|Zcold_del}) converges to that of cold gas metallicity for instantaneously produced elements in Eq.~(\ref{eq|Zcold_inst}). The behavior of $Z_{\rm cold}^\Delta$ for $\tau\simeq \Delta$ reads
\begin{equation}
Z_{\rm cold}^\Delta(\tau)\simeq \frac{s\,y_Z^\Delta\,(1-\mathcal{R})}{2}\, \frac{(s\gamma-1)\,e^{s\gamma\,\Delta/\tau_{\rm cond}}}{e^{(s\gamma-1)\,\Delta/\tau_{\rm cond}}-1} \left(\frac{\tau-\Delta}{\tau_{\rm cond}}\right)^2~.
\end{equation}

The above solution holds when a single delay time $\Delta$ is assumed. However, it is well known that type-I$a$ SNe feature a non-trivial delay time probability distributions ${\rm d}p/{\rm d}\Delta$, or DTDs.
Typically, universal DTDs with shapes ${\rm d}p/{\rm d}\Delta\propto \Delta^{-1}$, $\propto e^{-\Delta/\Delta_c}$, $\propto e^{-(\Delta-\bar\Delta)^2/\sigma_{\Delta}^2}$ or combination of these, are consistent with observations and widely adopted in chemical evolution models (e.g., Greggio 2005; Mannucci et al. 2006; Totani et al. 2008; Walcher et al. 2016; Schonrich \& Binney 2009; Maoz \& Graur 2017; for a review and further references, see Maoz et al. 2014). In such a case, one can easily recognize that, since the differential equation involved is linear, the above solution actually constitutes the Green function of the problem, i.e. the solution for a Dirac-$\delta_D$ delay time distribution centered at a delay time $\Delta$. Thus, the overall solution for a generic ${\rm d}p/{\rm d}\Delta$ is just the superposition of the previous solution for given $\Delta$ weighted by the DTD, so that
\begin{equation}
Z_{\rm cold}^{\rm DTD}(\tau) = \int{\rm d}\Delta\, \frac{{\rm d}p}{{\rm d}\Delta}\, Z_{\rm cold}^\Delta(\tau)~.
\end{equation}
As a working example, we take the analytically convenient exponential DTD with normalized shape ${\rm d}p/{\rm d}\Delta=(\omega/\tau_{\rm cond})\,e^{-\omega\,\Delta/\tau_{\rm cond}}$, where $\tau_{\rm cond}/\omega$ is  average (typical) DTD timescale. Then one can compute explicitly
\begin{empheq}[box=\fbox]{align}\label{eq|Zcold_DTD}
\nonumber\\
Z_{\rm cold}^{\rm DTD}(\tau) = \bar Z_{\rm cold}^{\rm DTD}\, \left[1-\left(\frac{s\gamma-1}{s\gamma-\omega}\right)^2\, \frac{e^{(s\,\gamma-\omega)\, x}-1-x\,(\omega-1)\,(s\gamma-\omega)/(s\gamma-1)}{ e^{(s\,\gamma-1)\, x}-1}\right]~,\\
\nonumber
\end{empheq}
where
\begin{equation}
\bar Z_{\rm cold}^{\rm DTD} = \frac{s\, \omega\,y_Z^\Delta\, (1-\mathcal{R})}{(\omega-1)\,(s\, \gamma-1)}~.
\end{equation}
As in Eqs.~(\ref{eq|basicsol}), one can easily check that for $\omega>0$ the above solution Eq.~(\ref{eq|Zcold_DTD}) is physically meaningful (non-negative for any $x$) when $s\gamma > 1$.
Note that the solution is also defined for $\omega\simeq 1$ and explicitly writes down as
\begin{equation}
Z_{\rm cold}^{\rm DTD}(\tau) \simeq s\, \,y_Z^\Delta\, (1-\mathcal{R})\, \left[\frac{x}{s\gamma-1}\,\frac{e^{(s\gamma-1)\,x}+1}{e^{(s\gamma-1)\,x}-1}-\frac{2}{(s\gamma-1)^2}\right]~.
\end{equation}

The early-time behavior of Eq.~(\ref{eq|Zcold_DTD}) for $\tau\ll \tau_{\rm cond}$ reads
\begin{equation}
Z_{\rm cold}^{\rm DTD}(\tau) \simeq \frac{s\, \omega\,y_Z^\Delta\, (1-\mathcal{R})}{6}\, \left(\frac{\tau}{\tau_{\rm cond}}\right)^2~,
\end{equation}
so that initially the increase in metallicity of delayed metals $Z_{\rm cold}^{\rm DTD}\propto \tau^2$, even when averaged over the DTD, is clearly retarded with respect to that of instantaneously produced metals $Z_{\rm cold}\propto \tau$. On the other hand, the late-time behavior for $\tau\gg \tau_{\rm cond}$ depends on $\omega$ as
\begin{equation}
\nonumber\\
Z_{\rm cold}^{\rm DTD}(\tau) \simeq \left\{
\begin{aligned}
&\frac{s\, \omega\,y_Z^\Delta\, (1-\mathcal{R})}{(\omega-1)\,(s\, \gamma-1)}
& {\rm for}~~~~\omega>1~,\\
\\
&\frac{s\, \,y_Z^\Delta\, (1-\mathcal{R})}{s\, \gamma-1}\, \frac{\tau}{\tau_{\rm cond}} & {\rm for}~~~~\omega=1 \\
\\
&\frac{s\, \omega\,y_Z^\Delta\, (1-\mathcal{R})\,(s\, \gamma-1)}{(1-\omega)\,(s\, \gamma-\omega)^2}\, e^{(1-\omega)\,\tau/\tau_{\rm cond}} & {\rm for}~~~~\omega<1
\end{aligned}
\right.\\
\nonumber
\end{equation}
so that the solution converges for $\omega>1$ to $\bar Z_{\rm cold}^{\rm DTD}$, while it diverges linearly for $\omega=1$ and exponentially for $\omega<1$; however, these divergences occur only at very late-times, so that the solution behaves very similarly
out to $\tau/\tau_{\rm cond}\lesssim 10^2$ for any value of $\omega$.

\subsubsection{Stellar metallicity}\label{sec|del_Z_star_subsect}

The metallicity $Z_{\rm \star}$ in the stellar component is computed by averaging the cold gas metallicity over the star formation history:
\begin{equation}\label{eq|Zstar_def}
Z_{\rm \star}(\tau) = \frac{1}{M_{\rm \star}(\tau)}\, \int_0^{\tau}{\rm d}\tau'~Z_{\rm cold}(\tau')\, \dot M_{\rm \star}(\tau')~,
\end{equation}
so that $Z_{\rm \star}$ represents the amount of metal stocked into the stellar component. Using the self-consistent expression of the cold gas metallicity for instantaneously produced elements, one obtains
\begin{empheq}[box=\fbox]{align}\label{eq|Zstar_inst}
\nonumber\\
Z_{\rm \star}(\tau) = \bar Z_{\rm \star}\, \left[1-\frac{s\, \gamma}{s\, \gamma-1}\, \frac{e^{-x}-e^{-s\, \gamma\, x}\, [1+(s\, \gamma-1)\, x]}{s\gamma-1+e^{-s\, \gamma\, x}-s\gamma\,e^{-x}}\right]~,\\
\nonumber
\end{empheq}
where the asymptotic behavior for $\tau\gg \tau_{\rm cond}$ writes
\begin{equation}
\bar Z_{\rm \star} = \frac{y_Z\,(1-\mathcal{R})}{\gamma} = \frac{s\gamma-1}{s\gamma}\, \bar Z_{\rm cold}~;
\end{equation}
it is seen that our analytic solutions, differently from other models in the literature, predict that the stellar metallicity is not equal, but rather somewhat lower, than the cold gas one. The early-time behavior of $Z_\star$ for $\tau\ll \tau_{\rm cond}$  reads
\begin{equation}
Z_{\rm \star}(\tau) \simeq \frac{s\,y_Z\, (1-\mathcal{R})}{3}\, \frac{\tau}{\tau_{\rm cond}}~,
\end{equation}
so that initially $Z_{\star}(\tau)\simeq 2\, Z_{\rm cold}(\tau)/3$, i.e., the stellar and cold gas metallicity evolve in parallel.

An analogous computation can be performed for the delayed cold gas metallicity (see Section \ref{sec|del_Z_gas_subsect}) by inserting $Z_{\rm cold}^\Delta$ from Eq.~(\ref{eq|Zcold_del}) in  Eq.~(\ref{eq|Zstar_def}), to yield
\begin{equation}\label{eq|Zstar_del}
Z_{\rm \star}^\Delta(\tau) = \left\{
\begin{aligned}
 \bar Z_{\rm \star}^\Delta\,& \left[1-\frac{s\, \gamma}{s\, \gamma-1}\, \frac{e^{-(x-x_\Delta)}-e^{-s\, \gamma\, (x-x_\Delta)}\, [1+(s\, \gamma-1)\, (x-x_\Delta)]}{s\gamma-1+e^{-s\, \gamma\, x}-s\gamma\, e^{-x}}+\right. \\
 & &\\
&-\left.
\frac{e^{-s\gamma\,x}-e^{-s\,\gamma\,(x-x_\Delta)}-s\gamma\,[e^{-x}-e^{-(x-x_\Delta)}]}{s\gamma-1+e^{-s\, \gamma\, x}-s\gamma\, e^{-x}}\right]~, & {\rm for}~~~~x\geq x_\Delta\\
& &\\
& 0 &  {\rm for}~~~~x<x_{\Delta}\\
& &
\end{aligned}
\right.
\end{equation}
where the value for $\tau\gg \tau_{\rm cond}$ reads
\begin{equation}
\bar Z_{\rm \star}^\Delta = \frac{y_Z^\Delta\,(1-\mathcal{R})}{\gamma}~.
\end{equation}
It is interesting to note that, if star formation proceeded for long times, the ratio of the stellar metallicity for instantaneously and delayed elements would amount to $\bar Z_\star^{\Delta=0}/\bar Z_\star^{\Delta}\approx y_Z/y_Z^{\Delta}$, i.e., the ratio of the respective yields. On the contrary, if star formation is quenched after some time (as it is the case for massive galaxies because of BH feedback, see Sect.~\ref{sec|feedbacks}) then the  different evolution of $Z_\star^{\Delta=0}$ in Eq.~(\ref{eq|Zstar_inst}) and of $Z_\star^{\Delta}$ in Eq.~(\ref{eq|Zstar_del}) would imply an underabundance of delayed with respect to instantaneously produced elements; this will be at the origin of the $\alpha$-enhancement (see Sect.~\ref{sec|starmetallicity_sect}).

Note that, as it should be, for $\Delta=0$ the time dependence in Eq.~(\ref{eq|Zstar_del}) converges to that of the stellar metallicity for instantaneously produced elements in Eq.~(\ref{eq|Zstar_inst}). The behavior of $Z_\star^\Delta$ for $\tau\simeq \Delta$ reads
\begin{equation}
Z_\star^\Delta(\tau)\simeq \frac{s\, y_Z^\Delta\, (1-\mathcal{R})}{6}\,\frac{s\gamma\,(s\gamma-1)}{s\gamma-1+e^{-s\gamma\,x_\Delta}-s\gamma\,e^{-x_\Delta}}\, \left(\frac{\tau-\Delta}{\tau_{\rm cond}}\right)^3~.
\end{equation}

In case of a non-trivial SN type-I$a$ delay time distribution DTD ${\rm d}p/{\rm d}\Delta$, the overall solution for the stellar metallicity may be derived from
\begin{equation}
Z_{\rm \star}^{\rm DTD}(\tau) = \int{\rm d}\Delta\, \frac{{\rm d}p}{{\rm d}\Delta}\, Z_{\rm \star}^\Delta(\tau) = \frac{1}{M_{\rm \star}(\tau)}\, \int_0^{\tau}{\rm d}\tau'~Z_{\rm cold}^{\rm DTD}(\tau')\, \dot M_{\rm \star}(\tau')~;
\end{equation}
the expression for an exponential DTD is still analytic but rather cumbersome, so we do not report it here.

\subsection{Dust}\label{sec|Dust}

We now turn to describe in simple analytic terms the global evolution of the dust mass and dust-to-gas mass ratio. As in many previous analytic approaches, we assume dust to consist of two interlinked components, namely, a refractory \emph{core} and a volatile \emph{mantle}, subject to the evolution equations
\begin{equation}\label{eq|dust}
\left\{
\begin{aligned}
\frac{\rm d}{{\rm d}\tau}(M_{\rm cold}\, D_{\rm core}) & =-\gamma\,\dot{M}_{\rm \star}\, D_{\rm core}-\kappa_{\rm SN}\,\dot{M}_{\rm \star}\, D_{\rm core}+{y_D}\,(1-\mathcal{R})\, \dot{M}_{\rm \star}~,\\
\\
\frac{\rm d}{{\rm d}\tau}(M_{\rm cold}\, D_{\rm mantle}) &=-\gamma\,\dot{M}_{\rm \star}\, D_{\rm mantle}-\kappa_{\rm SN}\,\dot{M}_{\star}\, D_{\rm mantle}+ \epsilon_{\rm acc}\, \dot{M}_{\rm \star}\, D_{\rm core}\,(Z-D_{\rm mantle})~.
\end{aligned}
\right.
\end{equation}
The first equation prescribes that the evolution of the mass in grain cores $M_{\rm cold}\,D_{\rm core}$ results from the competition of various processes: production due to stellar evolution at a rate $y_D\,(1-\mathcal{R})\,\dot{M}_{\rm \star}$ with an average yield $y_D$; astration by starformation and ejection from galactic outflows, that combine in the rate term $-\gamma\, \dot M_{\rm \star}\, D_{\rm core}$; dust sputtering, spallation and destruction via SN shockwaves at a rate $\kappa_{\rm SN}\, \dot{M}_{\rm \star}\, D_{\rm core}$ with a stength parameter $\kappa_{\rm SN}$. The second equation describes the evolution of the mass in dust mantles, which differs from the previous one for the production term: mantle growth is assumed to be driven by accretion of metals onto pre-existing grain cores at a rate $\epsilon_{\rm acc}\, \dot M_{\rm \star}\, D_{\rm core}\, (Z-D_{\rm mantle})$ with an efficiency $\epsilon_{\rm acc}$.

Using the self-consistent expression for the cold gas mass, Eqs.~(\ref{eq|dust}) can be recast in terms of the dust mass fractions:
\begin{equation}\label{mantle_eq}
\left\{
\begin{aligned}
\dot{D}_{\rm core}&= -\frac{D_{\rm core}}{\tau_{\rm cond}}\, \left[s\,\kappa_{\rm SN}+\frac{s\, \gamma-1}{1-e^{-(s\, \gamma-1)\, x}}\right]+\frac{y_D\, (1-\mathcal{R})\,s}{\tau_{\rm cond}}~;\\
 \\
\dot{D}_{\rm mantle}&= -\frac{D_{\rm mantle}}{\tau_{\rm cond}}\, \left[s\,\kappa_{\rm SN}+s\,\epsilon_{\rm acc}\, D_{\rm core}+\frac{s\, \gamma-1}{1-e^{-(s\, \gamma-1)\, x}}\right]+\frac{s\, \epsilon_{\rm acc}\, D_{\rm core}}{\tau_{\rm cond}}\, Z_{\rm cold}~.
\end{aligned}
\right.
\end{equation}
with initial conditions $D_{\rm core}(0)=D_{\rm mantle}(0)=0$. The corresponding analytic solution for grain cores is
\begin{empheq}[box=\fbox]{align}\label{eq|Dcore}
\nonumber\\
D_{\rm core}(\tau) = \bar D_{\rm core}\, \left[1-\frac{s\, \gamma-1}{e^{(s\, \gamma-1)\, x}-1}\, \frac{1-e^{-s\, \kappa_{\rm SN}\, x}}{s\, \kappa_{\rm SN}}\right]~,\\
\nonumber
\end{empheq}
where the asymptotic value for $\tau\gg \tau_{\rm cond}$ reads
\begin{equation}
\bar D_{\rm core} = \frac{s\,y_D\, (1-\mathcal{R})}{s\,(\gamma+\kappa_{\rm SN})-1}~.
\end{equation}

In solving the equation for the mantle, we assume the core fraction $D_{\rm core}(\tau)$ to be fixed at its asymptotic value $\bar D_{\rm core}\ll \bar Z_{\rm cold}$, since from Eq.~(\ref{eq|Dcore}) this is seen to be attained very rapidly after a time $\tau\ga \tau_{\rm cond}/s\, \kappa_{\rm SN}$; for typical values of the parameters (See Sect.~\ref{sec|otherparams} this amounts to a few $10^{-2}\, \tau_{\rm cond}$. We then obtain
\begin{empheq}[box=\fbox]{align}
\nonumber\\
D_{\rm mantle}(\tau) = \bar D_{\rm mantle}\, \left\{1-\frac{(s\, \gamma-1)\, x}{e^{(s\, \gamma-1)\, x}-1}\, \left[1+\frac{s\, \gamma-1}{ s\, \tilde\epsilon}\, \left(1-\frac{1-e^{-s\,\tilde\epsilon\, x}}{s\,\tilde \epsilon\, x}\right)\right]\right\}~,\\
\nonumber
\end{empheq}
where $\tilde\epsilon\equiv \kappa_{\rm SN}+\epsilon_{\rm acc}\,\bar D_{\rm core}$ and the asymptotic value for $\tau\gg \tau_{\rm cond}$ reads
\begin{equation}\label{mantle_asym_eq}
\bar D_{\rm mantle} = \frac{s\,\epsilon_{\rm acc}\, \bar D_{\rm core}\, \bar Z_{\rm cold}}{ s\,(\gamma+\kappa_{\rm SN}+\epsilon_{\rm acc}\, \bar D_{\rm core})-1}~;
\end{equation}
thus if the accretion process is very efficient, Eq.~(\ref{mantle_asym_eq}) implies that the final dust fraction tends to the gas metallicity.
The early-time behavior for $\tau\ll \tau_{\rm cond}$ writes
\begin{equation}
\left\{
\begin{aligned}
D_{\rm core}(\tau) &\simeq \frac{s\,y_D\,(1-\mathcal{R})}{2}\,\frac{\tau}{ \tau_{\rm cond}}~,\\
\\
D_{\rm mantle}(\tau) &\simeq \frac{s^3}{6}\, \frac{\epsilon_{\rm acc}\, y_D\, y_Z\,(1-\mathcal{R})^2}{s\, (\gamma+\kappa_{\rm SN})-1}\, \left(\frac{\tau}{\tau_{\rm cond}}\right)^2~,
\end{aligned}
\right.
\end{equation}
so that the mantle component overwhelms the core one soon after dust production has started. Note that the second equation above is strictly valid for times $\tau_{\rm cond}/s\kappa_{\rm SN}\la \tau\la \tau_{\rm cond}$ when $D_{\rm core}$ has already saturated to its asymptotic value $\bar D_{\rm core}$.

It is worth stressing that total dust production and enrichment (core plus mantle) are in general very rapid with respect to the condensation timescale, of order a few $10^{-1}\, \tau_{\rm cond}$; this will be extremely relevant for very high redshift starforming galaxies and quasar hosts (see Sect.~\ref{sec|dustmass}).

\subsection{Outflowing mass and metals}\label{sec|FedOut}

The effect of galactic outflows driven by SN explosions and stellar winds is to heat and remove the cold gas at a rate $\epsilon_{\rm out}\, \dot M_\star$ proportional to the star formation rate. One can also include a sort of impulsive energy/momentum feedback; e.g., this may be driven by emission from the central supermassive black hole, that in the progenitors of elliptical galaxies is thought to quench star formation. To a crude approximation, the action of such impulsive feedback can be described as heating/ejection of all the residual mass in cold gas $M_{\rm cold}(\tau_{\rm burst})$ at a galactic age $\tau_{\rm burst}$. Then the overall outflowing gas mass at $\tau_{\rm burst}$ reads
\begin{equation}\label{M_out_eq}
M_{\rm out} = \epsilon_{\rm out}\, M_{\star}(\tau_{\rm burst}) + M_{\rm cold}(\tau_{\rm burst})~.
\end{equation}

In addition, the metallicity of the mass outflow by SNe and stellar winds is $\epsilon_{\rm out}\,Z_{\star}$, while the one associated to impulsive feedback is $Z_{\rm cold}(\tau_{\rm burst})$. Thus, the average metallicity of the outflowing gas at $\tau_{\rm burst}$ writes
\begin{equation}\label{Z_out_eq}
Z_{\rm out} = \frac{\epsilon_{\rm out}\, Z_{\star}(\tau_{\rm burst})\, M_{\star}(\tau_{\rm burst}) + Z_{\rm cold}(\tau_{\rm burst})\, M_{\rm cold}(\tau_{\rm burst})}{\epsilon_{\rm out}\, M_{\star}(\tau_{\rm burst}) + M_{\rm cold}(\tau_{\rm burst})}~,
\end{equation}
and is found to strike an intermediate course between the cold gas and the stellar metallicity. This will be relevant for the enrichment of the warm and hot medium pervading/surrounding massive ETGs (see Sect.~\ref{sec|outmetals}).

\section{Application to ETGs and their starforming progenitors}\label{sec|Application}

We now specialize the analytic solutions presented in the previous sections to a particular issue: the formation of ETGs and the evolution of their high-$z$ starforming progenitors. To this purpose, in Sect.~\ref{sec|biascoll}-\ref{sec|feedbacks}-\ref{sec|otherparams} we set the parameters entering our analytic solutions via physical arguments inspired by in-situ galaxy-BH coevolution scenarios for ETG  formation (e.g., Granato et al. 2004; Lapi et al. 2006, 2014, 2018), with particular focus on the role of cooling/condensation and feedback processes; then in Sect.~\ref{sec|merging_sect} we discuss the mass additions by mergers in the DM and stellar components, and in Sect.~\ref{sec|average_sect} we describe how to deal with different formation redshifts in order to obtain observable spatially-averaged quantities.

\subsection{Infall fraction and starformation timescales}\label{sec|biascoll}

To start with, we set the infall fraction $f_{\rm inf}$, condensation timescale $\tau_{\rm cond}$ and  the ratio between the star formation to condensation timescale $s=\tau_{\rm cond}/\tau_{\star}$ entering the analytic solutions (in particular the boxed equations in Sect.~\ref{sec|An_sol}). We base on the in-situ galaxy-BH coevolution scenario by Lapi et al. (2018); we recall here the basic notions relevant for the present context, and defer the reader that paper for an extended discussion.

The in-situ scenario envisages that only a fraction $f_{\rm inf}=M_{\rm inf}/f_b\, M_{\rm H}$ of the available baryons $f_b\, M_{\rm H}$ in a halo of mass $M_{\rm H}$, initially located within a radius $R_{\rm inf}$, is able to cool and fall in toward the central region of the galaxy where strong star formation takes place; the radius $R_{\rm inf}$ and the infall fraction can be estimated along the following lines. Given a halo of mass $M_{\rm H}$ formed at redshift $z_{\rm form}$, its virial radius and virial circular velocity are approximately given by
$R_{\rm H}\approx 110\, M_{\rm H,12}^{1/3}$ $\,[E_{z_{\rm form}}/E_{z_{\rm form}=2}]^{-1/3}$ kpc and $v_{c, \rm H} \approx 200\, M_{\rm H,12}\, [E_{z_{\rm form}}/E_{z_{\rm form}=2}]^{1/6}$ km s$^{-1}$ in terms of the redshift dependent factor $E_{z_{\rm form}}=\Omega_\Lambda+\Omega_M\,(1+z_{\rm form})^3$ and the normalized halo mass $M_{\rm H,12}=M_{\rm H}/10^{12}\, M_\odot$. Adopting a standard NFW (Navarro et al. 1997) profile for the DM component yields an approximate scaling\footnote{For a NFW profile, the logarithmic slope of the mass distribution $M_{\rm H}(< r)\propto r^{\mu}$ reads $\mu\equiv {\rm d}\log M_{\rm H}/{\rm d} \log r = [c\,x/(1+c\,x)]^2\,[\ln(1+c\,x)-c\,x\,(1+c\,x)]^{-1}$ in terms of
the normalized radius $x\equiv r/R_{\rm H}$ and of the concentration parameter $c$. For a concentration $c\approx 4$ typical of massive galaxy halos formed at $z_{\rm form}\approx 2$ (e.g., Bullock et al. 2001; Zhao et al. 2003), the slope $\mu$ takes on values from $0.8$ to $1.2$ in moving from $R_{\rm H}$ to $0.3\, R_{\rm H}$, and can be effectively approximated with unity down to $\sim 0.4-0.6\, R_{\rm H}$. For smaller radii, the slope progressively approaches the central value $\mu\sim 2$, which can be approximately used for $r\la 0.1\, R_{\rm H}$.} $M_{\rm H}(<r)\propto r$, so that $R_{\rm inf}\simeq f_{\rm inf}\, R_{\rm H}$ and $M_{\rm H}(<R_{\rm inf})\simeq f_{\rm inf}\,M_{\rm H}(<R_{\rm H})$; therefore the dynamical time at $R_{\rm inf}$ is given by
\begin{equation}
t_{\rm dyn}(R_{\rm inf})\simeq\frac{\pi}{2}\sqrt{\frac{R_{\rm inf}^3}{G\,M_{\rm H}(<R_{\rm inf})}}\approx 6\times 10^8\,f_{\rm inf}\,[E_{z_{\rm form}}/E_{z_{\rm form}=2}]^{-1/2}\, \rm{yr}~.
\end{equation}

On the other hand, the cooling time at $R_{\rm inf}$ reads
\begin{equation}
t_{\rm cool}(R_{\rm inf})\simeq
\frac{3\,k_{\rm B}T}{2\,\mu\, \mathcal{C}\,n(R_{\rm inf})\,\Lambda(T,Z)}
\end{equation}
where $T$ is the temperature, $\mu\approx 0.6$ is the mean molecular weight, $n(R_{\rm inf})$ is the gas density, $\mathcal{C}$ is the clumping factor and $\Lambda(T,\,Z)$ is the cooling function in cgs units dependent on temperature and metallicity (e.g., Sutherland \& Dopita 1993). The infalling gas is expected to have temperatures close to the virial $T\simeq 0.5\,\mu\,m_p\,v_{c,\rm H}^2\approx 1.5\times 10^6\, M_{\rm H,12}^{2/3}\,[E_{z_{\rm form}}/E_{z_{\rm form}=2}]^{1/3}$ K; correspondingly, $\Lambda(T,Z)\ga 2\times 10^{-23}$ erg cm$^3$ s$^{-1}$ for $Z\ga Z_\odot/10$ (recall from Sect.~\ref{sec|Met} that this value is attained quite rapidly within $\la 10^{-1}\, \tau_{\rm cond}\sim $ a few $10^7$ yr). The gas density is expected to be on the order of the average baryon density within $R_{\rm inf}$, which reads $n(R_{\rm inf})\approx 4 \times 10^{-4}\, f_{\rm inf}^{-2}\,[E_{z_{\rm form}}/E_{z_{\rm form}=2}]$ cm$^{-3}$ having assumed $n(r)$ to follow an isothermal distribution; the clumping factor is expected to be close to that in the IGM, which cosmological simulations (see Iliev et al. 2007; Pawlik et al. 2009; Finlator et al. 2012; Shull et al. 2012) indicate around $\mathcal{C}\sim 6-20$ at $z\simeq 2$, so that we take $\mathcal{C}\approx 10$.

\begin{figure}[!ht]
  \centering
\includegraphics[width=.9\textwidth]{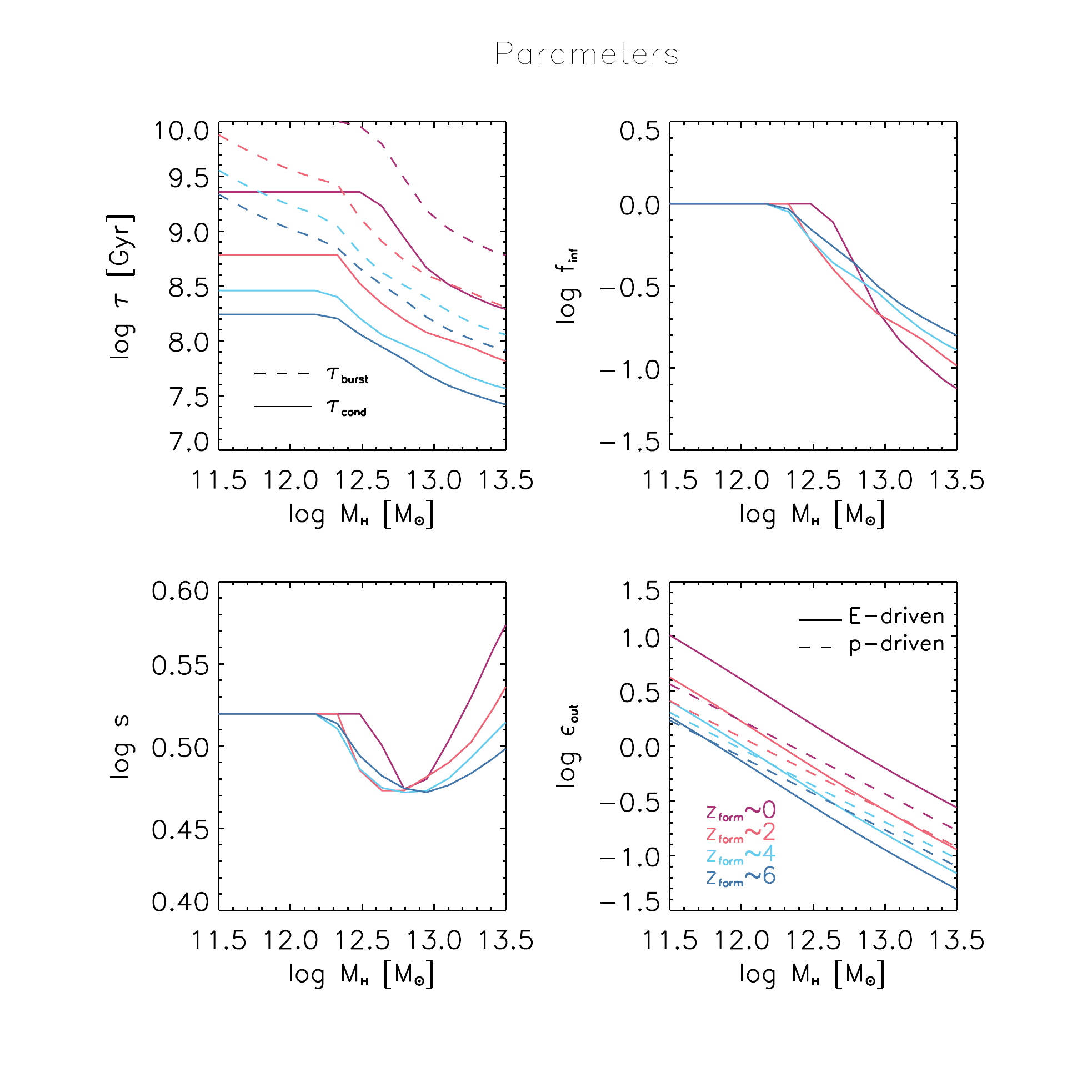}
\caption{Main parameters entering the analytic solutions relevant for ETG progenitors, as a function of host halo mass $M_{\rm H}$ and formation redshift $z_{\rm form}$. Top left panel: condensation timescale $\tau_{\rm cond}$ (solid lines) and duration of the starforming phase $\tau_{\rm burst}$ (dashed lines). Top right panel: infall fraction $f_{\rm inf}$. Bottom left panel: ratio $s\equiv \tau_{\rm cond}/\tau_{\rm \star}$ of the starformation to the condensation timescale. Bottom right panel: mass loading factor of the outflows from stellar feedback $\epsilon_{\rm out}$, for energy-driven (solid lines) and momentum-driven (dashed lines) outflows. In all panels the color-code refer to different formation redshifts $z_{\rm form}=0$ (red), $2$ (orange), $4$ (cyan), and $6$ (blue).}\label{fig|params}
\end{figure}

When $t_{\rm dyn}\la t_{\rm cool}$ the gas can cool efficiently and infall toward the central regions on the dynamical timescale; therefore we set $f_{\rm inf}$ by requiring that the above dynamical $t_{\rm dyn}(R_{\rm inf})\propto f_{\rm inf}$ and cooling time $t_{\rm cool}\propto f_{\rm inf}^2$ match. Then we prescribe that the fraction $f_{\rm inf}$ of the baryons located within $R_{\rm inf}$ can cool and condense toward the central regions over a timescale $\tau_{\rm cond}\approx t_{\rm dyn}(R_{\rm inf})$. We plot $f_{\rm inf}$ and $\tau_{\rm cond}$ as a function of the halo mass and formation redshift in Fig.~\ref{fig|params}. The parameter $f_{\rm inf}$ is essentially unity for halo masses $M_{\rm H}\la$ a few $10^{12}\, M_\odot$, while for larger masses it drops to low values because cooling becomes progressively inefficient and prevents condensation toward the central regions; the dependence on formation redshift is negligible. As to $\tau_{\rm cond}$, it scales with halo mass similarly to $f_{\rm inf}$, while the constant value for $M_{\rm H}\la$ a few $10^{12}\, M_\odot$ depends on formation redshift approximately as $(1+z_{\rm form})^{-3/2}$, reflecting the increased density of the ambient medium at earlier cosmic epochs.

To proceed further, note that such infalling gas rotates, being endowed with the specific angular momentum
\begin{equation}\label{eq|jhalo}
j_{\rm inf} \equiv f_{\rm inf}\,j_{\rm H}\approx 1100\, f_{\rm inf}\,\lambda_{0.035}\,M_{\rm H,12}^{2/3}\, [E_{z_{\rm form}}/E_{z_{\rm form}=2}]^{-1/6}~{\rm km~s^{-1}~kpc}~;
\end{equation}
here $j_{\rm H} \equiv \sqrt{2}\, \lambda\, R_{\rm H}\, v_{c, \rm H}$ is the halo specific angular momentum and $\lambda_{0.035}\equiv \lambda/0.035$ is the halo spin parameter.
Numerical simulations (see Barnes \& Efstathiou 1987; Bullock et al. 2001; Macci\'o et al. 2007; Zjupa \& Springel 2017) have shown that $\lambda$ exhibits a log-normal distribution with average value $\langle\lambda\rangle\approx 0.035$ and dispersion $\sigma_{\log \lambda} \approx 0.25$ dex, nearly independent of mass and redshift. In deriving the above equation the specific angular momentum of the baryonic gas has been assumed to initially follow the radial profile of the DM's $j_{\rm H}(<r)\propto M^\alpha_{\rm H}(<r)$ with $\alpha\approx 1$, which in turn is found from simulations to closely follow the mass profile (e.g., Bullock et al. 2001; Shi et al. 2017).

Given that $j_{\rm inf}$ is approximately conserved, the inflow of the gas from $R_{\rm inf}$ toward the central region can proceed until the radius $R_{\rm rot}$ where the rotational support balance the gravitational pull, i.e., $G\, M_{\rm tot}(<R_{\rm rot})/R_{\rm rot}^2=j_{\rm inf}^2/R_{\rm rot}^3$; such a condition yields
\begin{equation}\label{eq|Rrot}
R_{\rm rot}\simeq \frac{j_{\rm inf}^2}{G\, M_{\rm inf}}\approx 1.5\, \lambda_{0.035}^2\,f_{\rm inf}\, [E_{z_{\rm form}}/E_{z_{\rm form}=2}]^{-1/3}~~{\rm kpc}~,
\end{equation}
and the corresponding dynamical time amounts to
\begin{equation}\label{eq|tdyn_Rrot}
t_{\rm dyn}(R_{\rm rot}) \simeq \frac{\pi}{2}\,\sqrt{\frac{R^3_{\rm rot}}{G\, M_{\rm inf}}}\approx 4\times 10^6\,\lambda_{0.035}^{3}\,f_{\rm inf}\, [E_{z_{\rm form}}/E_{z_{\rm form}=2}]^{-1/2}~~{\rm yr}~.
\end{equation}
Recent observations of high redshift starforming galaxies have revealed that most of the SFR occur
within a compact region a few kiloparsecs under heavy enshrouded conditions (e.g., Scoville et al. 2014, 2016; Ikarashi et al. 2015; Simpson et al. 2015; Straatman et al. 2015; Spilker et al. 2016; Tadaki et al. 2017a,b); in particular, the size $R_{\rm rot}$ derived above has been shown by Lapi et al. (2018) to be consistent with those measured via far-IR/submillimeter and CO line observations of $z\sim1-2$ starforming galaxies (e.g., Barro et al. 2016, 2017; Hodge et al. 2016; Tadaki et al. 2017a; Talia et al. 2018; Lang et al. 2019). Further infall of the gas within $R_{\rm rot}$ can only occur by spreading out specific angular momentum via dynamical friction and gravitational torques, as indicated by specific simulations and suggested by dynamical measurements in high-redshift starforming galaxies (see Dekel et al. 2009; Genzel et al. 2011; Zolotov et al. 2015; for a review, Bournaud 2016 and references therein); eventually these processes will cause a significant smearing of the initial correlation between the baryon and halo specific angular momenta, especially in the high-redshift progenitors of ETGs where angular momentum loss is expected to drive formation of a substantial bulge component (e.g., Danovich et al. 2015; Jiang et al. 2019).

The cold baryonic gas mass $M_{\rm cold}(R_{\rm rot})\la f_{\rm inf}\, f_b\, M_{\rm H}\sim 10^{10}-10^{11}\, M_\odot$, namely a fraction $f_{\rm inf}$ of the baryons already present in the halo at formation, will be driven by such processes within $R_{\rm rot}\sim$ kpc and will form star on a timescale $t_{\rm SFR}(R_{\rm rot})\approx 50\times t_{\rm dyn}(R_{\rm rot})\approx (1-2)\times 10^8$ yr (see Elmegreen et al. 2005; Krumholz et al. 2012 and references therein), implying SFRs $\dot M_\star\la M_{\rm cold}(R_{\rm rot})/t_{\rm SFR}\la 10^2-10^3\, M_\odot$ yr$^{-1}$ and hence rapid metallicity enrichment and dust formation. These conditions have been indeed observed in ETG progenitors, in particular via high-resolution interferometric observations with ALMA (see Scoville et al. 2016; Barro et al. 2017; Tadaki et al. 2017a,b; Talia et al. 2018; Lang et al. 2019). Note that in such systems cooling and the ensuing star formation from baryons accreted during later halo growth over cosmological timescales will be hampered by the effects of stellar and AGN feedbacks (see below); this is at variance with respect to present-day disc-dominated galaxies, where the cold gas accretion is not inhibited by strong feedbacks and the star formation process is prolonged over several Gyrs. On the basis of the above discussion, we set the parameter $s\equiv \tau_{\rm cond}/\tau_\star$ entering our analytic solutions as $s=t_{\rm dyn}(R_{\rm inf})/t_{\rm SFR}(R_{\rm rot})$; in Fig.~\ref{fig|params} it is shown to take on values around $3-3.5$, weakly dependent on halo mass and formation redshift.

\subsection{Feedback parameters}\label{sec|feedbacks}

Another important parameter entering our analytic solutions is the mass loading factor $\epsilon_{\rm out}$ of slow outflows related to type-II SNe and stellar winds. We adopt the standard expression used in many literature studies since the pioneering work of White \& Frenk (1991; for reviews and further references see Mo et al. 2010 and Somerville \& Dave 2015)
\begin{equation}\label{eq|epsout}
\epsilon_{\rm out} = \frac{\epsilon_{\rm SN}\,\eta_{\rm SN}\, E_{\rm SN}}{E_{\rm bind}} \approx \epsilon_{SN,0.05}\, \eta_{\rm SN,-2}\, E_{\rm SN,51}\, M_{\rm H,12}^{-2/3}\, [E_{z_{\rm form}}/E_{z_{\rm form}=2}]^{-1/3}~,
\end{equation}
in terms of the occurrence $\eta_{\rm SN,-2}=\eta_{\rm SN}/0.01\, M_\odot$ of SNe per unit solar mass formed into stars (apt for our fiducial Chabrier IMF), of the average energy $E_{\rm SN, 51}=E_{\rm SN}/10^{51}$ erg per single SN explosion, of the energy fraction $\epsilon_{\rm SN,0.05}=\epsilon_{\rm SN}/0.05$ effectively coupled to the interstellar medium and available to drive the outflow (comparable to the overall coupled energy when cooperative SN blastwave propagation takes place, like in violent starbursts; e.g., Mac Low \& Ferrara 1999; Mo et al. 2010), and of the gas specific binding energy $E_{\rm bind}\approx 2.5\times 10^{14}\, M_{\rm H,12}^{2/3}\, [E_{z_{\rm form}}/E_{z_{\rm form}=2}]^{1/3}$ cm$^2$ s$^{-2}$ in the halo potential well (see Zhao et al. 2003; Mo \& Mao 2004). The outcome is illustrated in Fig.~\ref{fig|params} as a function of halo mass and formation redshift; it takes on values in the range $\epsilon_{\rm out}\sim 0.1-10$ for halo masses $M_{\rm H}\sim 10^{13.5}-10^{11}\, M_\odot$, with a weak dependence on formation redshift. Such a behavior is indeed in agreement with self-consistent hydrodynamical simulations of stellar feedback (e.g., Hopkins et al. 2012).

Note that the above Eq.~(\ref{eq|epsout})  is meant to describe an energy-driven, slow outflow that can offset gas infall from large scales out to $R_{\rm inf}\gg R_{\rm rot}$, where the binding energy is dominated by the DM halo. An alternative prescription for stellar feedback invokes momentum-driven outflows (e.g., Murray et al. 2005; Oppenheimer \& Dave 2006), that in our context can indeed operate to blow some of the cold gas out of $R_{\rm rot}$, where the binding energy is largely dominated by baryons; in such a case the relevant mass loading factor reads $\epsilon_{\rm out}= \eta_{\rm SN}\, E_{\rm SN}/v_{\rm SN}\, v_{\rm rot}\approx v_{\rm SN,3}\,  \eta_{\rm SN,-2}\, E_{\rm SN,51}\, \lambda_{0.035}\, M_{\rm H,12}^{-2/3}\,[E_{z_{\rm form}}/E_{z_{\rm form}=2}]^{-1/6}$ in terms of the typical velocity of SN ejecta $v_{\rm SN,3}\equiv v_{\rm SN}/10^3$ km s$^{-1}$ and of the escape velocity $v_{\rm rot}\approx j_{\rm inf}/R_{\rm rot}$ at $R_{\rm rot}$. The resulting $\epsilon_{\rm out}$ for the energy-driven and momentum-driven outflows are comparable within the uncertainties of the parameters entering the respective expressions (cf. Fig.~\ref{fig|params}), and produce similar evolution in the gas/stellar mass and metallicity when used in our analytic solutions.

Another form of feedback, that is thought to be extremely relevant in the formation of ETGs, is related to the hosted accreting supermassive BH (see e.g., Silk \& Rees 1998; Fabian 1999; King 2003; Granato et al. 2004; Murray et al. 2005; Lapi et al. 2006, 2014; for a review, see King 2014 and references therein). In this paper we do not aim at building up a self-consistent model of the coevolution between a starforming galaxy and the central supermassive BH, but we will describe phenomenologically the impact of BH feedback on the host galaxy by abruptly quenching the star formation and ejecting the residual gas mass after a time $\tau_{\rm burst}$. Constraints on $\tau_{\rm burst}$ comes from SED modeling of dusty ETG starforming progenitors (see Papovich et al. 2011; Moustakas et al. 2013; Steinhardt et al. 2014; da Cunha et al. 2015; Citro et al. 2016; Cassara et al. 2016; Boquien et al. 2019), that suggest a duration $\la 0.5-1$ Gyr for massive galaxies with $M_\star\ga$ a few $10^{10}\, M_\odot$. Similar values in massive ETGs are also concurrently indicated by local observations the stellar mass-metallicity relationship and of the $\alpha$-enhancement, i.e., iron under-abundance compared to $\alpha$-elements (occuring because star formation is stopped before type-I$a$ SN explosions can pollute the interstellar medium with substantial amounts of iron; e.g., Thomas et al. 2005; Gallazzi et al. 2006, 2014; for a review, see Renzini 2006). On the contrary, in low-mass spheroidal galaxies with $M_{\star}\la 10^{10}\, M_\odot$ data on the age of the stellar population and chemical abundances indicate that star formation proceeded for longer times, mainly regulated by SN feedback and stellar winds (see review by Conroy 2013). On this basis, we phenomenologically parameterize the timescale for the duration of the main SFR episode via the smooth expression
\begin{equation}\label{eq|tau_burst}
\tau_{\rm burst}\simeq 3\,\tau_{\rm cond}\,(1+M_{\rm H,12}^{-1})
\end{equation}
holding in the range $M_{\rm H}\sim 10^{11}-10^{13.5}\, M_\odot$. This formula is meant to interpolate between the aforementioned behaviors for less and more massive galaxies; e.g., at $z_{\rm form}\approx 2$ it prescribes short timescales $\tau_{\rm burst}\sim $a few $10^{8}$ yr for galaxies in massive halos with $M_{\rm H}\sim 10^{13}\, M_\odot$, and appreciably longer $\tau_{\rm burst}\sim$ several $10^{9}$  for galaxies in small halos with $M_{\rm H}\sim 10^{11.5}\, M_\odot$. We have checked that our results are rather insensitive to the specific shape of such a phenomenological parameterization. With a similar formula Mancuso et al. (2016a,b) have reproduced the main sequence of starforming galaxies at $z\approx 2$, and Lapi et al. (2017) have linked the statistics of starforming galaxies, AGNs and massive quiescent galaxies via a continuity equation approach.

Thus infall, condensation and feedback processes actually sets the main parameters entering our analytic solutions.

\subsection{Yields and other IMF-related parameters}\label{sec|otherparams}

Other relevant parameters are mainly determined by the adopted Chabrier (2003, 2005) IMF and the Romano et al. (2010) stellar yield models. We compute an instantaneous recycling fraction $\mathcal{R}\approx 0.45$, an average yield of instantaneously produced metals $y_Z\approx 0.06$ and an yield of oxygen $y_O\approx 0.04$ (see also Krumholz et al. 2012; Feldmann 2015; Vincenzo et al. 2016). Note that these values are weakly dependent on the chemical composition and somewhat dependent on the stellar yield models (e.g., Romano et al. 2010; Nomoto et al. 2013; Vincenzo et al. 2016); e.g., the metal yield can vary within the range $y_Z\sim 0.05-0.07$.

The yield for delayed metals is adopted to be $y_Z^\Delta\approx 2.7\times 10^{-3}$; this has been computed taking into account that the occurrence of type-I$a$ SNe is around $2\times 10^{-3}/M_\odot$ per stellar mass formed, and that $0.63\,M_\odot$ of iron-group elements are produced on average per explosion (see Bell et al. 2003; Maoz et al. 2014); the outcome is also consistent with the normalization of the observed type-I$a$ SN delay time distribution (e.g., Vincenzo et al. 2017).

Finally, in the treatment of dust production, we adopt a dust yield $y_D\approx 7\times 10^{-4}$ (see Bianchi \& Schneider 2007; Zhukovska et al. 2008; Feldmann 2015), a strength parameter $\kappa_{\rm SN}\approx 10$ for dust spallation by SN winds (see McKee 1989; de Bennassuti et al. 2014), and an efficiency for dust accretion $\epsilon_{\rm acc}\approx 10^{5}$ (see Hirashita 2000; Asano et al. 2013; Feldmann 2015); these parameters are rather uncertain, and mainly set basing on previous literature and on obtaining a good match to the dust vs. stellar mass relationship observed in $z\ga 2$ starforming galaxies.

\subsection{Halo and stellar mass growth by mergers}\label{sec|merging_sect}

After virialization, both the DM halo and the stellar content are expected to grow because of merging events. To describe these processes, we rely on the outcomes of $N-$body and hydro simulations, and in particular the \texttt{Illustris} project (see \url{http://www.illustris-project.org/}).
As to the halo growth, the merger rates per bin of redshift $z$ and of halo mass ratio $\mu_{\rm H}$ can be described with the fitting formula originally proposed by Fakhouri \& Ma (2008), Fakhouri et al. (2010) and Lapi et al. (2013)
\begin{equation}
\frac{{\rm d}N_{\rm H, merg}}{{\rm d}t\, {\rm d}\mu_{\rm H}} = N_{\rm H}\, M_{\rm H,12}^a\, \mu_{\rm H}^{-b-2}\, e^{(\mu_{\rm H}/\tilde\mu_{\rm H})^c}\, \frac{{\rm d}\delta_c}{{\rm d}t}
\end{equation}
in terms of the descendant halo mass $M_{\rm H,12}=M_{\rm H}/10^{12}\, M_\odot$, and of the linear threshold for collapse $\delta_c(z)$.
Genel et al. (2010) have determined the parameters entering the above expression from the \texttt{Illustris}-Dark simulations, finding $N_{\rm H}=0.065$, $a=0.15$, $b=-0.3$, $c=0.5$ and $\tilde\mu_{\rm H}=0.4$. The average halo mass growth $\langle\dot M_{\rm H\, merg}\rangle$ is obtained by multiplying the above expression by $\mu_{\rm H}/(1+\mu_{\rm H})$ and integrating over $\mu_{\rm H}$ from a minimum value $\mu_{\rm H, min}$; we use $\mu_{\rm H, min}\approx 10^{-5}$ that corresponds to include all mergers, since the contribution from smaller mass ratios to the halo mass growth is essentially negligible.

\begin{figure}[!ht]
  \centering
\includegraphics[width=1\textwidth]{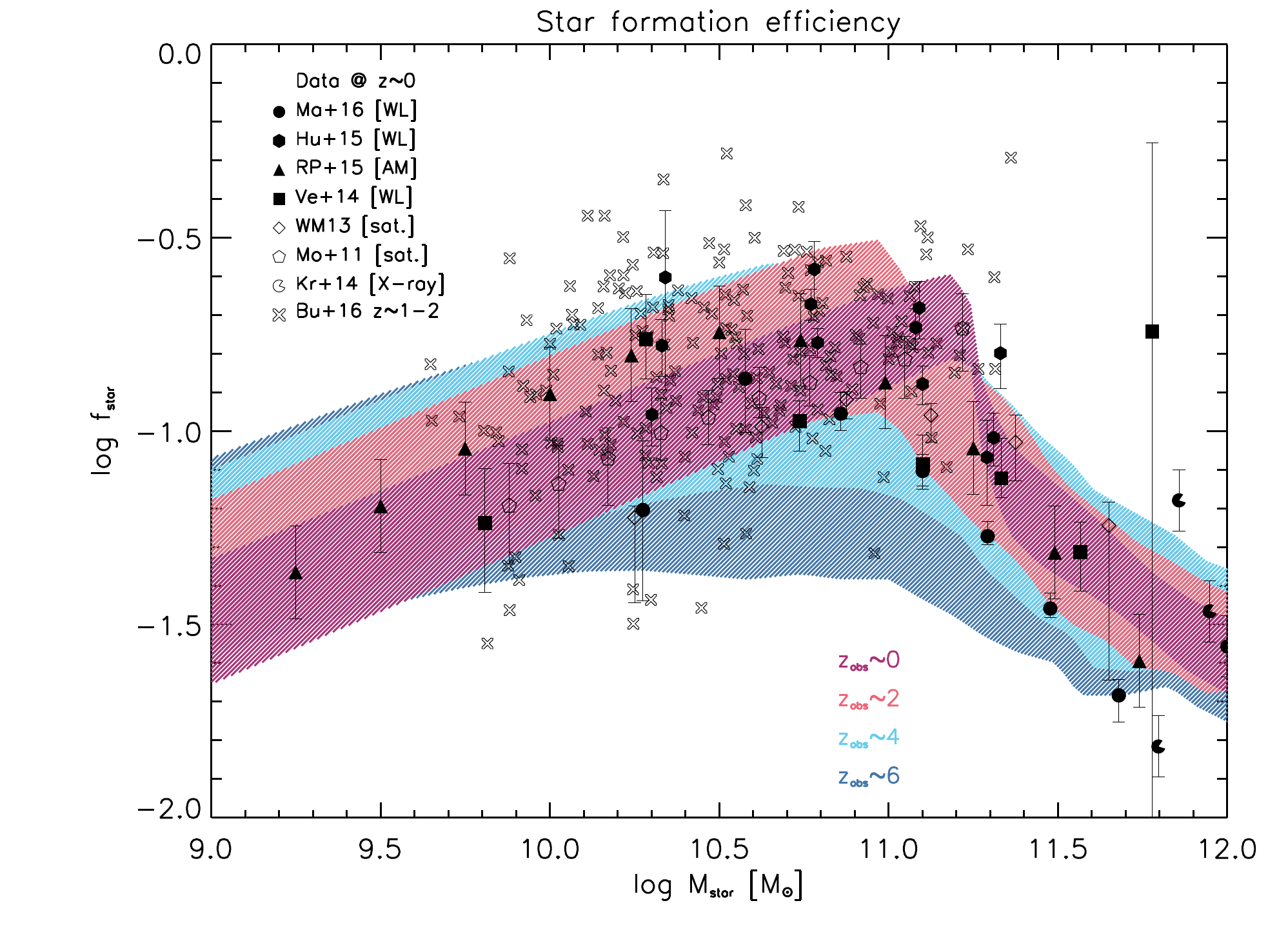}
\caption{Star formation efficiency $f_{\star}=M_{\star}/f_b\,M_H$ vs. stellar mass $M_{\star}$, at different observation redshifts $z\approx 0$ (red), $2$ (orange), $4$ (cyan), and $6$ (blue); the shaded areas illustrate the $1\sigma$ variance associated to the average over different formation redshifts.
Data points are from Mandelbaum et al. (2016; circles), Hudson et al. (2015; hexagons) and Velander et al. (2014; squares) via weak lensing, Rodriguez-Puebla et al. (2015; triangles) via subhalo abundance matching, Wojtak \& Mamon (2013; diamonds) and More et al. (2011; pentagons) via satellite kinematics, Kravtsov et al. (2014, pacmans) via X-ray observations of BCGs, and Burkert et al. (2016; crosses) via mass profile modeling of galaxies at $z\sim 1-2$. If not indicated explicitly, error bars are $\approx 0.25$ dex.}\label{fig|fstar}
\end{figure}

As to the stellar mass growth, Rodriguez-Gomez et al. (2015, 2016)
have inferred from the analysis of the full hydrodynamic \texttt{Illustris} simulation the fitting formula
\begin{equation}
\frac{{\rm d}N_{\star,\rm merg}}{{\rm d}t\, {\rm d}\mu_\star} = N_{\star}(z_{\rm form})\, M_{\star,10}^{a(z_{\rm form})}\, [1+(M_{\star}/\tilde M_{\star})^{d(z_{\rm form})}]\, \mu_\star^{b(z_{\rm form})+c\log M_{\star,10}}
\end{equation}
in terms of the descendant stellar mass $M_{\star,10}=M_{\star}/10^{10}\, M_\odot$, where $\tilde M_{\star}=2\times 10^{11}\, M_\odot$, $N_{\star}(z_{\rm form})=N_0\, (1+z_{\rm form})^{N_1}$ with $\log N_0 [{\rm  Gyr}^{-1}]=-2.2287$ and $N_1=2.4644$, $a(z_{\rm form})=a_0\, (1+z_{\rm form})^{a_1}$ with $a_0=0.2241$ and $a_1=-1.1759$, $b(z_{\rm form})=b_0\, (1+z_{\rm form})^{b_1}$ with $b_0=-1.2595$ and $b_1=0.0611$, $c=-0.0477$, $d(z_{\rm form})=d_0\, (1+z_{\rm form})^{d_1}$ with $d_0=0.7668$ and $d_1=-0.4695$. To obtain the average stellar mass growth $\langle\dot M_{\star,\rm merg}\rangle$ via mergers we follow Rodriguez-Gomez et al. (2016) in multiplying the above expression by $\mu_\star/(1+3\, \mu_\star)$ and integrating over $\mu_\star$ from a minimum value $\mu_{\star \rm min}\approx 10^{-2}$; the latter value practically corresponds to include all relevant mergers, since the cumulative effect from those with smaller stellar mass ratios is negligible with respect to the stellar mass growth (see Rodriguez-Gomez et al. 2016). Note that we exploit the above merging rates from \textsl{Illustris} but not the related prescriptions for in-situ star formation, that are instead based on our framework; this approach is aimed to reproduce the halo to stellar mass ratio at different redshifts (see Sect.~\ref{sec|fstar_sect} and Fig.~\ref{fig|fstar}).

\begin{figure}[!ht]
\centering
\includegraphics[width=.8\textwidth]{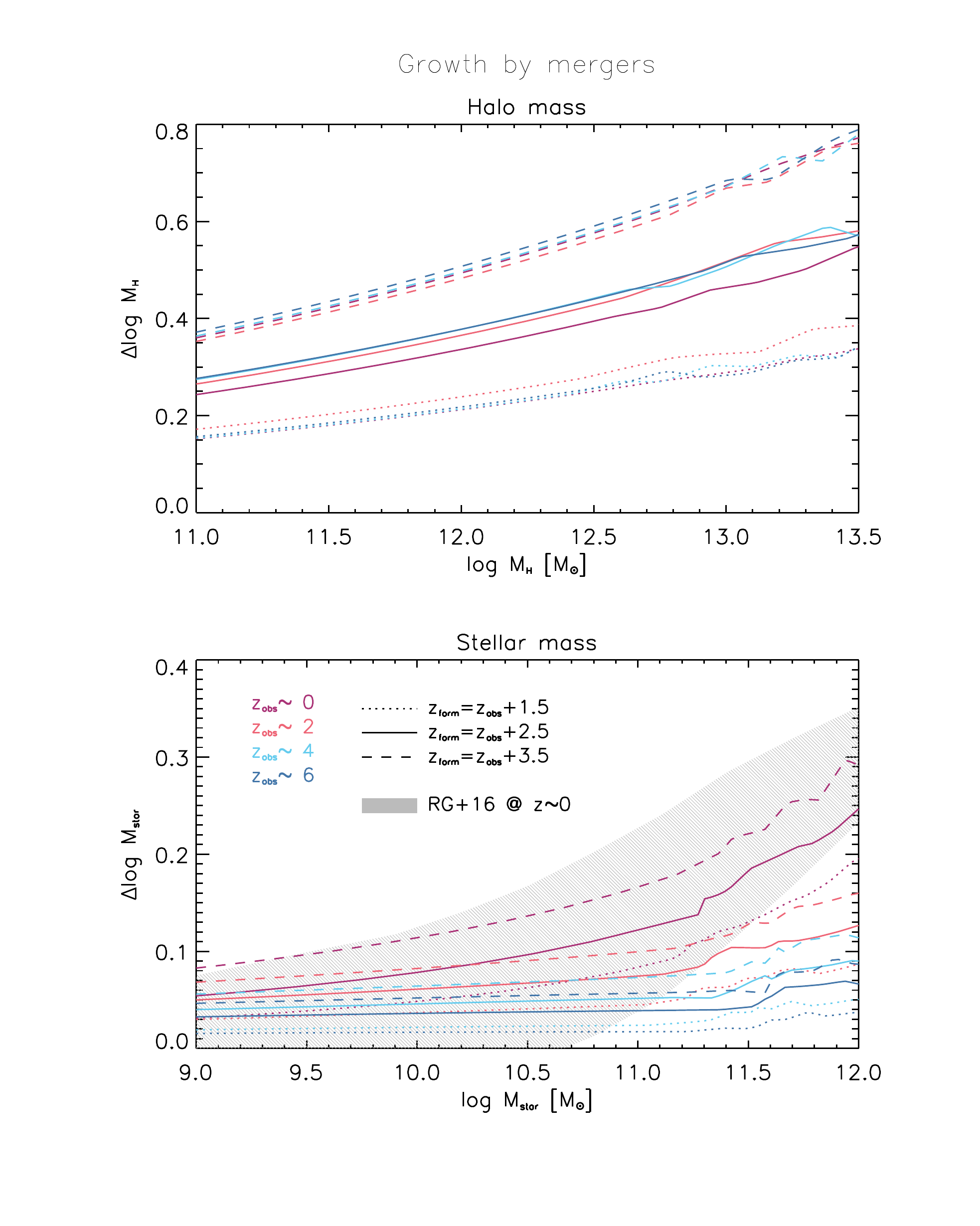}
\caption{Growth of the halo mass (top panel) and of the stellar mass (bottom panel) by mergers, as a function of the final halo and stellar masses, respectively. We illustrate the outcomes at observation redshifts $z_{\rm obs}=0$ (red), $2$ (orange), $4$ (cyan), and $6$ (blue) for different formation redshifts $z_{\rm form}=z_{\rm obs}+1.5$ (dotted), $z_{\rm form}=z_{\rm obs}+2.5$ (solid) and $z_{\rm form}=z_{\rm obs}+3.5$ (dashed). For $z_{\rm obs}=0$ we also show the typical stellar mass growth via mergers from the \textsl{Illustris} simulation by Rodriguez-Gomez et al. (2016; grey shaded area).}\label{fig|merging}
\end{figure}

During each timestep ${\rm d}t$ after halo formation at $z_{\rm form}$ till the observation redshift $z_{\rm obs}$, the halo and the stellar mass are increased by the amounts $\langle \dot M_{\rm merg,H}\rangle\, {\rm d}t$ and $\langle \dot M_{\rm merg,star}\rangle\, {\rm d}t$, respectively. To have a grasp on such mass additions, in Fig.~\ref{fig|merging} we illustrate the growth of halo and stellar mass via mergers as a function of the descendant final masses. Specifically, we show the outcomes at observation redshifts $z_{\rm obs} = 0$, $2$, $4$, and $6$ (color coded), and different formation redshifts $z_{\rm form}=z_{\rm obs}+1.5$, $z_{\rm obs}+2.5$, and $z_{\rm obs}+3.5$ (linestyle coded). Plainly, at given observation and formation redshift the amount of relative mass addition by mergers increases with the descendant mass, while at given descendant mass and observation redshift the mass additions increase for increasing formation redshift; in addition, the halo mass addition depends weakly on the observation redshift, while the stellar mass addition appreciably increases for decreasing $z_{\rm obs}$. Typically, the halo mass increase at $z_{\rm obs}\approx 0$ for a descendant halo mass $M_{\rm H}\approx 10^{13}\, M_\odot$ formed at $z_{\rm form}\approx 2.5$ amounts to a factor $\sim 3$; correspondingly, the stellar mass increase at $z_{\rm obs}\approx 0$ for a galaxy with descendant stellar mass $M_\star\approx 3\times 10^{11}\, M_\odot$ formed at $z_{\rm form}\approx 2.5$ amounts to $\sim 50\%$.

\subsection{Average over formation redshift}\label{sec|average_sect}

In order to derive the statistical properties of the galaxy population concerning a quantity $\mathcal{Q}$ (e.g., halo mass, star formation rate, stellar mass), we proceed as follows. We exploit the analytic solutions of Sect.~\ref{sec|An_sol} for the evolution of individual galaxies with different formation redshift $z_{\rm form}$ and halo masses $M_{\rm H}$ at formation (the halo mass at $z_{\rm obs}$ is larger because of mass addition by mergers, see previous Sect.), to obtain $\mathcal{Q}(\tau|M_{\rm H},z_{\rm form})$. Given an observation redshift $z_{\rm obs}\la z_{\rm form}$, we pick up the value of $\mathcal{Q}$ at $\tau=t_{z_{\rm obs}}-t_{z_{\rm form}}$, where $t_z$ is the cosmic time at redshift $z$. Finally, we perform the average over different formation redshift to get
\begin{equation}\label{eq|zformave}
\langle \mathcal{Q}\rangle(M_{\rm H},z_{\rm obs})\propto \int^{\infty}_{z_{\rm obs}}{\rm d}z_{\rm form}\, \frac{{\rm d}^2N_{\rm H}}{{\rm d}\log M_{\rm H}\, {\rm d}z_{\rm form}}\, \mathcal{Q}(t_{z_{\rm obs}}-t_{z_{\rm form}}|M_{\rm H},z_{\rm form})
\end{equation}
where ${\rm d}^2N_{\rm H}/{\rm d}\log M_{\rm H}\, {\rm d}z_{\rm form}$ is the halo formation rate computed via the excursion set framework, and checked against $N-$body suimulations, by Lapi et al. (2013; see also Lacey \& Cole 1993; Kitayama \& Suto 1996; Moreno et al. 2009; Giocoli et al. 2012). The normalization constant in the above Eq.~(\ref{eq|zformave}) is clearly the same integral without $\mathcal{Q}$, and the $1\sigma$ variance is computed as $\sigma_{\mathcal{Q}}=\sqrt{\langle \mathcal{Q}^2\rangle-\langle \mathcal{Q}\rangle^2}$.

To compute the formation rate one needs the halo mass function, i.e., the co-moving number density of halo per halo mass bin, as provided by state-of-the-art N-body simulations; we adopt the determination by Tinker et al. (2008; see also Watson et al. 2013; Bocquet et al. 2016; Comparat et al. 2017, 2019). Since we are mainly concerned with the properties of ETGs residing at the center of halos, we actually exploit the galaxy halo mass function, i.e., the mass function of halos hosting one individual galaxy (though the difference with respect to the halo mass function emerge only for $z\la 1$ and $M_{\rm H}\ga$ several $10^{13}\, M_\odot$). This can be built up from the overall halo mass function by adding to it the contribution of sub-halos and by probabilistically removing from it the contribution of halos corresponding to galaxy systems via halo occupation distribution modeling; we refer the reader to Appendix A of Aversa et al. (2015) for details on such a procedure.

\section{Results}\label{sec|Results}

We now present, compare with recent data, and discuss a bunch of results from our analytic solutions applied to ETGs and their starforming progenitors.

\subsection{Time evolution of individual galaxies}\label{sec|timevo}

\begin{figure}[!ht]
  \centering
\includegraphics[width=.9\textwidth]{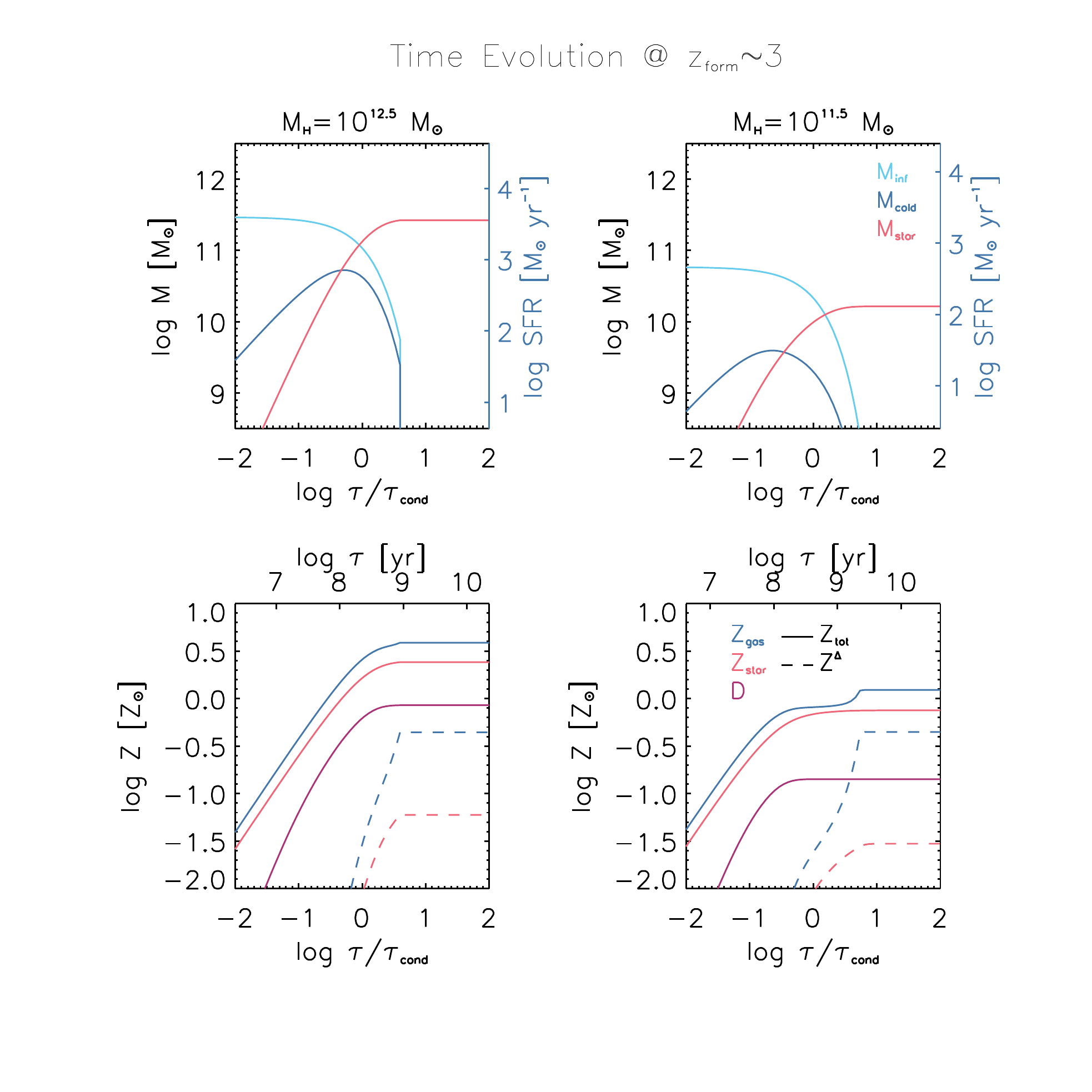}
\caption{Evolution of the mass components (top panels), metallicity and dust-to-gas ratio (bottom panels) as a function of the galactic age (normalized to the condensation timescale $\tau_{\rm cond}$ or in absolute units of yr), for galaxies hosted in halos with mass $M_{\rm H}=10^{12.5}\, M_\odot$ (left panels) and $M_{\rm H}=10^{11.5}\, M_\odot$ (right panels) at formation redshift $z_{\rm form}=3$; to highlight in-situ evolution, halo and stellar mass additions by mergers are switched off. In the top panels, cyan lines refer to the infalling mass $M_{\rm inf}$, orange lines to the stellar mass $M_{\star}$ (actually the integral of the SFR), blue lines to the cold gas mass $M_{\rm cold}$ (the corresponding SFR values $\dot M_\star=s\,M_{\rm cold}/\tau_{\rm cond}$ can be read off on the right y-axis). In the bottom panels, blue lines refer to the cold gas metallicity $Z_{\rm cold}$, orange lines to the stellar metallicity $Z_{\star}$, and red lines to the dust-to-gas mass ratio $D$; moreover solid lines are for the total metallicity, while dashed lines highlights the contribution from delayed metals.}\label{fig|timevo_z3}
\end{figure}

\begin{figure}[!ht]
  \centering
\includegraphics[width=.9\textwidth]{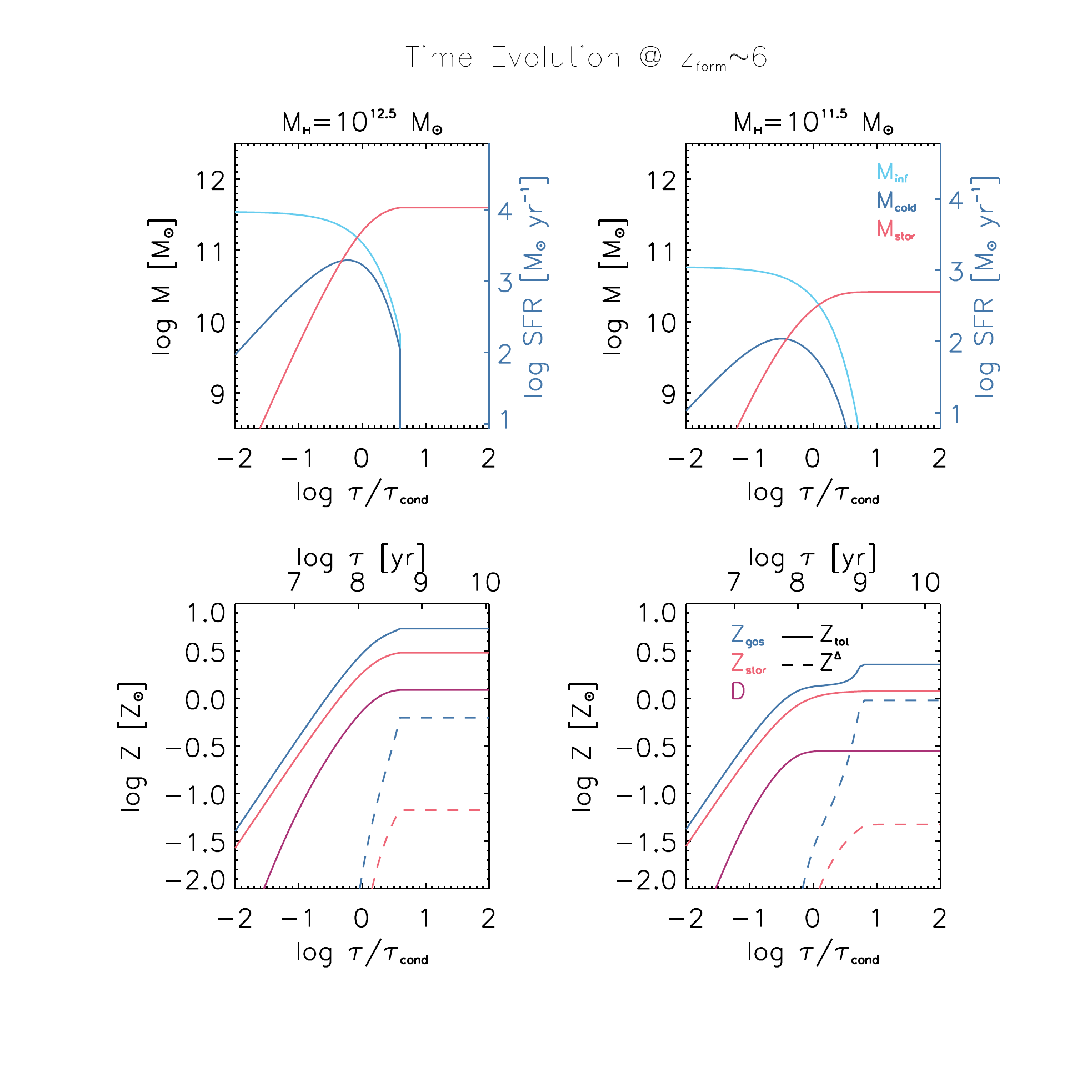}
\caption{Same as previous figure at formation redshift $z_{\rm form}=6$.}\label{fig|timevo_z6}
\end{figure}

We start by presenting the evolution with galactic age $\tau$ of the relevant spatially-averaged quantities described by our analytic solutions: infalling gas mass, cold gas mass, stellar mass, gas and stellar metallicity, dust mass. In Figs.~\ref{fig|timevo_z3} and \ref{fig|timevo_z6} we illustrate representative galaxies with different halo masses $M_{\rm H}=10^{12.5}\,M_\odot$ and $10^{11.5}\, M_\odot$ formed at different redshifts $z_{\rm form}=3$ and $6$.

The infalling gas mass decreases exponentially with the galactic age as it condenses in the cold gas component, which in turn feeds star formation; on the other hand, both components are affected by SN feedback. The balance of these processes makes the cold gas mass (hence the SFR) to slowly grow at early times, to attain a maximum and then to decrease exponentially; correspondingly, the stellar component increases almost linearly and then saturates. In massive halos with $M_{\rm H}\ga 10^{12}\,M_\odot$ the star formation is abruptly quenched by BH feedback, soon after $\tau_{\rm burst}\sim $ a few $10^8$ yr, while the residual gas mass is removed; conversely, in less massive halos $M_{\rm H}\la 10^{12}\,M_\odot$ BH feedback is inactive, so that the star formation can proceed for longer times $\tau_{\rm burst}\ga $ a few $10^9$ yr till the gas reservoir gets exhausted. This differential behavior is induced by the dependence of $\tau_{\rm burst}$ on halo mass, as specified by Eq.~(\ref{eq|tau_burst}), and phenomenologically renders the well-known downsizing behavior. In galaxy halos formed earlier, the typical SFRs are larger, and the accumulated stellar masses higher.

The metallicity in cold gas and in stars rises almost linearly, so that after a few $10^7$ yr it attains values above $\ga Z_\odot/10$; note that this rapid increase is particularly relevant for the metal enrichment of high redshift galaxies and QSO hosts (see Maiolino et al. 2005; Wang et al. 2008; Omont et al. 2013; Willott et al. 2015; Michalowski 2015; Venemans et al. 2018). In massive halos the metallicity saturates after a time a few $10^8$ yr to (super)solar values, and the contribution by delayed metals from type-I$a$ SNe to the stellar metallicity is minor since the BH feedback stops the star formation before any significant pollution. On the other hand, in low mass halos only subsolar values are attained, and the contribution from delayed metals becomes relatively more important, especially at old ages; this differential behavior in the relative fraction of instantaneous and delayed metals will be at the origin of the observed $\alpha$-enhancement in massive galaxies. The dust-to-gas ratio follows a behavior similar to the metal enrichment, with a quite rapid growth within $\sim 10^8$ yr driven mainly by accretion onto preformed core grains, up to a saturation value which can be close to unity in massive halos, while it stays $\la 0.3$ in smaller ones. We stress that the evolution of metals and dust is quite rapid, and being mainly related to in-situ processes, is weakly dependent on redshift; this could be important for detecting primordial galaxies at $z\ga 6$ in future wide-area IR surveys, that will be routinely achievable with ALMA and with the advent of the JWST (see De Rossi \& Bromm 2019).

\subsection{Star formation efficiency}\label{sec|fstar_sect}

In Fig.~\ref{fig|fstar} we show the star formation efficiency $f_{\star}\equiv M_{\star}/f_b\, M_{\rm §H}$, namely the fraction of initial baryonic mass converted into stars, as a function of the stellar mass for different observation redshifts $z_{\rm obs}=0$, $2$, $4$, and $6$ (color-coded); the shaded areas illustrate the 1$\sigma$ scatter associated to the average over formation redshifts.
The star formation efficiency $f_{\star}$ is found to be a non-monotonic function of the stellar mass $M_{\star}$, with a maximum value of $20-30\%$ slowly increasing with redshift around $M_{\star}\simeq 10^{11}\, M_\odot$, and a decrease to values less than $10\%$ for $M_{\star}\sim$ a few 10$^9\, M_\odot$ and for $M_{\star}\simeq 10^{12}\, M_\odot$; all in all, star formation in galaxies is a very inefficient process.

Such a behavior is easily understood in terms of infall/condensation and feedback processes. At small masses,
infall and condensation are efficient ($f_{\rm inf}\approx 1$) but star formation is regulated
by outflows from SNe and stellar winds; conversely, at high masses infall and condensation
become less efficient ($f_{\rm inf}\la 1$) and star formation is also hindered by
BH feedback. All in all,
the maximum value of the star formation efficiency occurs at a mass  corresponding approximately to the transition between the stellar and BH feedback (see Shankar et al. 2006;
Moster et al. 2013; Aversa et al. 2015).

Our result at $z_{\rm obs}\approx 0$ is compared with the local data for
ETGs from various authors, determined
via weak lensing (see Velander et al. 2014; Hudson et al. 2015;
Rodriguez-Puebla et al. 2015; Mandelbaum et al. 2016) and
satellite kinematics (see More et al. 2011; Wojtak \&
Mamon 2013), while the outcome at $z\approx 2$ should be compared with the
estimates by Burkert et al. (2016) via H$\alpha$ data and mass profile modeling;
we find very good agreement in normalization and scatter
within the still large observational uncertainties.
Our results as a function of redshift are also similar, within a factor of $2$, to the
determinations via abundance matching technique by Moster et al.
(2013), Behroozi et al. (2013), Aversa et al. (2015), and Lapi et al. (2017).

We stress that the similarity of the efficiency at $z\approx 2$ to the local values is
indicative that star formation is mainly an in situ process (see
Lilly et al. 2013; Moster et al. 2013; Aversa et al. 2015; Mancuso et al. 2016a); actually,
the difference in $f_\star$ between these redshifts is mainly driven by the increase in the stellar
and halo mass due to late-time mergers.

\subsection{Galaxy main-sequence}\label{sec|MS_sect}

\begin{figure}[!ht]
\centering
\includegraphics[width=1\textwidth]{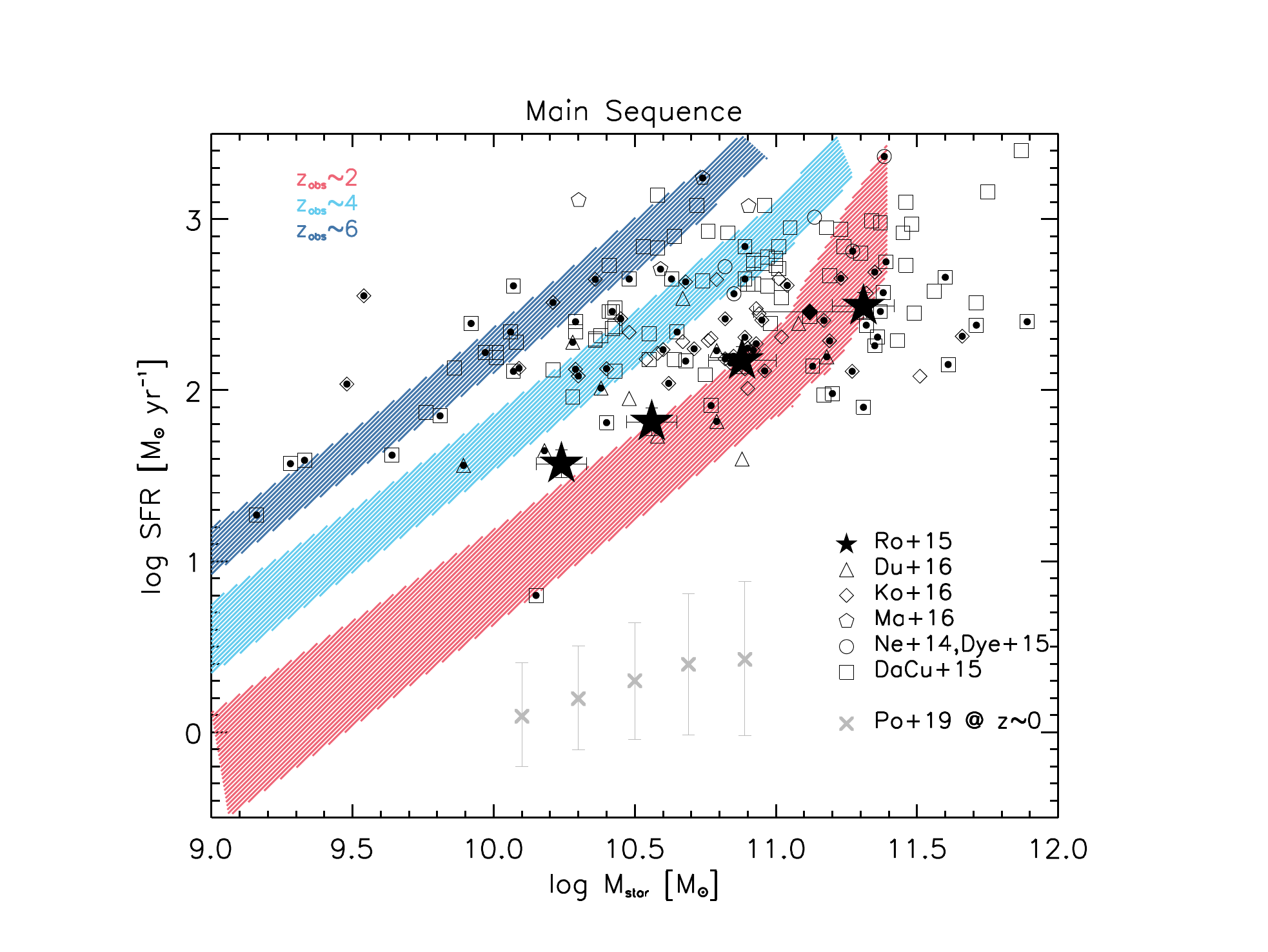}
\caption{Star formation rate SFR vs. stellar mass $M_{\star}$, alias the main sequence of starforming galaxies, at different observation redshifts $z\approx 2$ (orange), $4$ (cyan), and $6$ (blue); the shaded areas illustrate the $1\sigma$ variance associated to the average over different formation redshifts.
The black filled stars are the observational determinations of
the main sequence at $z\sim 2$ based on the statistics of large mass-selected samples by Rodighiero et al. (2015). The other symbols (error bars omitted for clarity) refer to far-IR data for individual objects at $z\sim 1-4$ (those in the range $z\sim 1.5-2.5$ are marked with a dot) by Dunlop et al. (2017; triangles), Koprowski et al. (2016; diamonds), Ma et al. (2016; pentagons), Negrello et al. (2014), Dye et al. (2015; circles), and da Cunha et al. (2015; squares); for reference the determination at $z\approx 0$ by Popesso et al. (2019; crosses) is also reported.}\label{fig|MS}
\end{figure}

In Fig.~\ref{fig|MS} we present our results concerning the so-called {\itshape main-sequence} (MS) of starforming galaxies (e.g., Elbaz et al. 2007; Daddi et al. 2007; Noeske et al. 2007; Rodighiero et al. 2011, 2015;
Speagle et al. 2014; Whitaker et al. 2014; Schreiber et al. 2017; Tacchella et al. 2018b; Popesso et al. 2019), namely the relation between the SFR and stellar mass at different observation redshifts $z_{\rm obs}= 2$, $4$, $6$ (color-coded).
The outcomes at $z\sim 2$ are in pleasing agreement with the observational determination from the large
statistics of mass-selected galaxy samples by Rodighiero et al. (2015); this further substantiates our solutions for the time evolution of the star formation and stellar mass in individual galaxies.

To highlight the relevance of observational selections different from that based on stellar mass, in Fig.~\ref{fig|MS} we also report data points for individual, far-IR-selected galaxies by Koprowski et al. (2016), Ma et al. (2016), Negrello et al. (2014), along with Dye et al. (2015), da Cunha et al. (2015), and Dunlop et al. (2017), mainly at redshifts $z\sim  1-4$. An appreciable fraction of the individual, far-IR selected galaxies around $z\sim 2$ (highlighted by dots within hollow symbols) lie above the main sequence, i.e., at SFR values higher than expected on the basis of the average relationship at a given $M_{\rm \star}$. These off-main-sequence objects can be simply interpreted (see Mancuso et al. 2016b; Lapi et al. 2017) as galaxies caught in an early evolutionary stage and still accumulating their stellar mass. Thus, young starforming
galaxies are found to be preferentially located above the main sequence or, better, to the left of it. As time goes by and the stellar mass increases, the galaxy moves toward the average main-sequence relationship, around which it will spend most of its lifetime. Afterwards, the SFR is quenched by feedback and the galaxy will then evolve passively to become a local early type; it will then populate a region of the SFR versus stellar mass diagram that is substantially below the main sequence. These loci of “red and dead” galaxies are indeed observed locally (see Renzini \& Peng 2015) and start to be pinpointed even at increasing redshift (see Man et al. 2016). For reference in Fig.~\ref{fig|MS} we also report the determination at $z\approx 0$ by Popesso et al. (2019), which is mainly representative of the local disk-dominated galaxies (not included in our framework); typically, these have star formation histories prolonged over several Gyrs and continue to form stars, though at rather low rates of a few $M_\odot$ yr$^{-1}$, even toward $z\approx 0$.

The overall redshift evolution of the main sequence for ETG and their progenitors is consistent with a scenario which traces the bulk of the star formation in galaxies back to local, in situ condensation processes. Specifically, at higher $z$ and in massive galaxies, the interstellar medium is on average denser and the condensation/star formation timescales are shorter. Thus the star formation in a galaxy of given stellar mass is higher, causing the main sequence locus to shift upwards. We stress that, moving toward higher redshift, the fraction of off-main-sequence objects
decreases appreciably; this is because, given the evolution of the SFR function and the shorter age of the universe, it is more and more difficult to spot galaxies of appreciably different ages and featuring very high SFRs.

\subsection{Gas mass}\label{sec|gasmass_sect}

\begin{figure}[!ht]
  \centering
\includegraphics[width=1\textwidth]{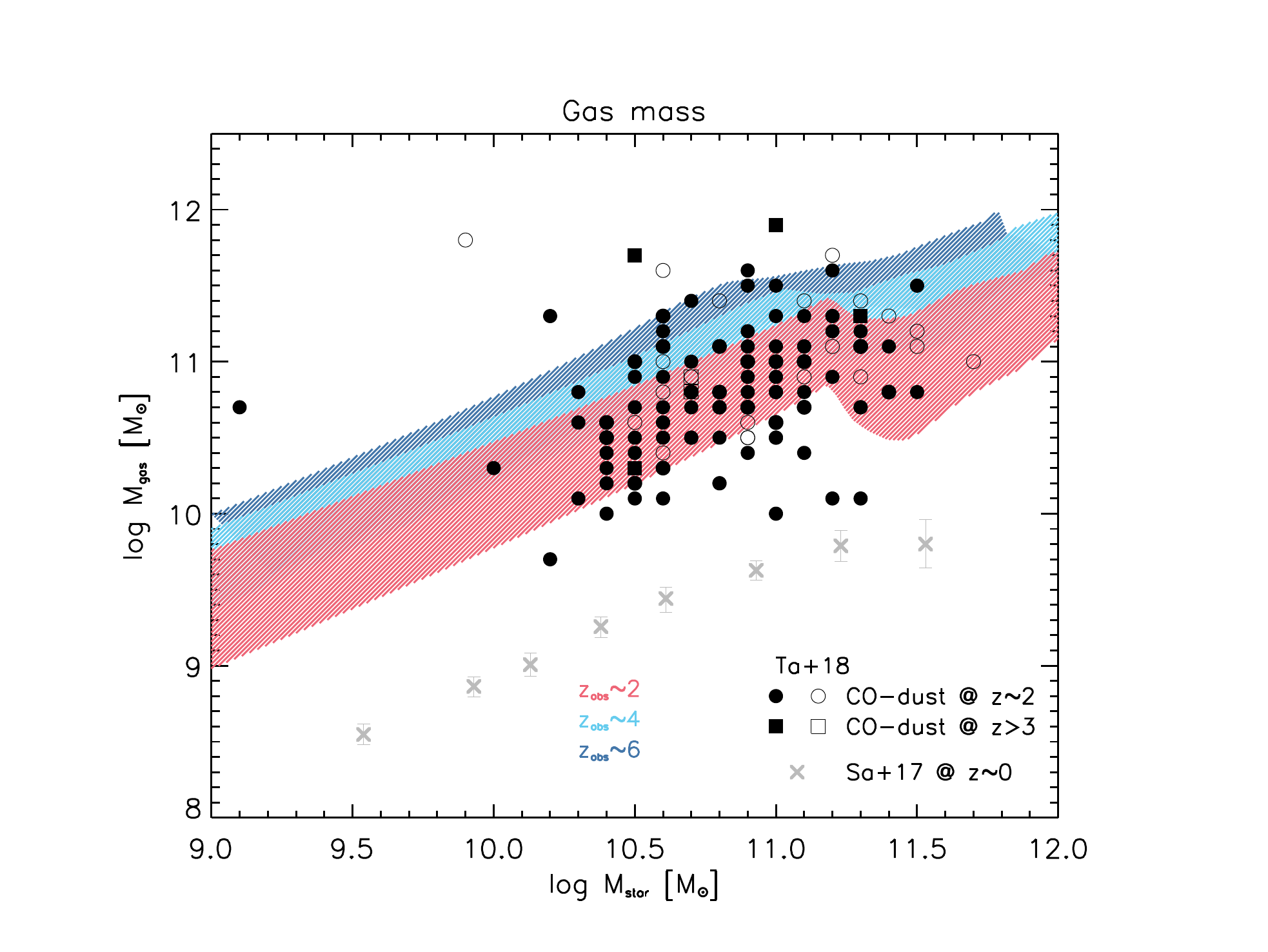}
\caption{Gas mass $M_{\rm gas}$ vs. stellar mass $M_{\star}$ at different observation redshifts, $z\approx 2$ (orange), $4$ (cyan), and $6$ (blue); the shaded areas illustrate the $1\sigma$ variance associated to the average over different formation redshifts. Data points are from Tacconi et al. (2018); circles represent objects at $z\sim 2$ while squares stand for galaxies at  redshifts $z\ga 3$. Filled symbols refer to gas mass estimates from CO lines, and empty symbols from dust far-infrared/sub-mm continuum. Error bars on datapoints (omitted for clarity) are of order $\approx 0.25$ dex; for reference data at $z\approx 0$ from Saintonge et al. (2017; crosses) are also reported.}\label{fig|gasmass}
\end{figure}

In Fig.~\ref{fig|gasmass} we illustrate the relationship between the gas mass $M_{\rm gas}$ and the stellar mass $M_{\star}$ for different observation redshifts $z_{\rm obs}\sim 2$, $4$ and $6$ (color-coded). Over most of the stellar mass range, the gas mass increases monotonically; the relationship holds up to $M_{\star}\sim 10^{11}\, M_\odot$, where a decrease in the gas mass is enforced since the infall/condensation processes becomes inefficient. As to the redshift evolution, the gas mass is  slightly higher for starforming galaxies observed at earlier epochs; this can be traced back to the fact that objects observed at higher redshift are on average younger and are expected to have converted less gas into stars. Note that at $z_{\rm obs}=0$ the descendants of these starforming galaxies, i.e., local ETGs, have very small gas reservoirs, since most of the gas has been consumed via star formation and/or ejected by feedback events. In fact, the $z\approx 0$ determination from Saintonge et al. (2017), reported as a reference, refers mainly to disk-dominated galaxies which are gas rich and still star-forming in the local Universe.

Our results for starforming ETG progenitors are compared with the gas mass estimates from Tacconi et al. (2018) from ALMA data for a large sample of starforming galaxies at redshifts $z\sim 1-4$; the agreement with the data is good in normalization, scatter and redshift evolution, although still large observational uncertainties hinder a more detailed comparison.

\subsection{Dust mass}\label{sec|dustmass}

\begin{figure}[!ht]
  \centering
\includegraphics[width=1\textwidth]{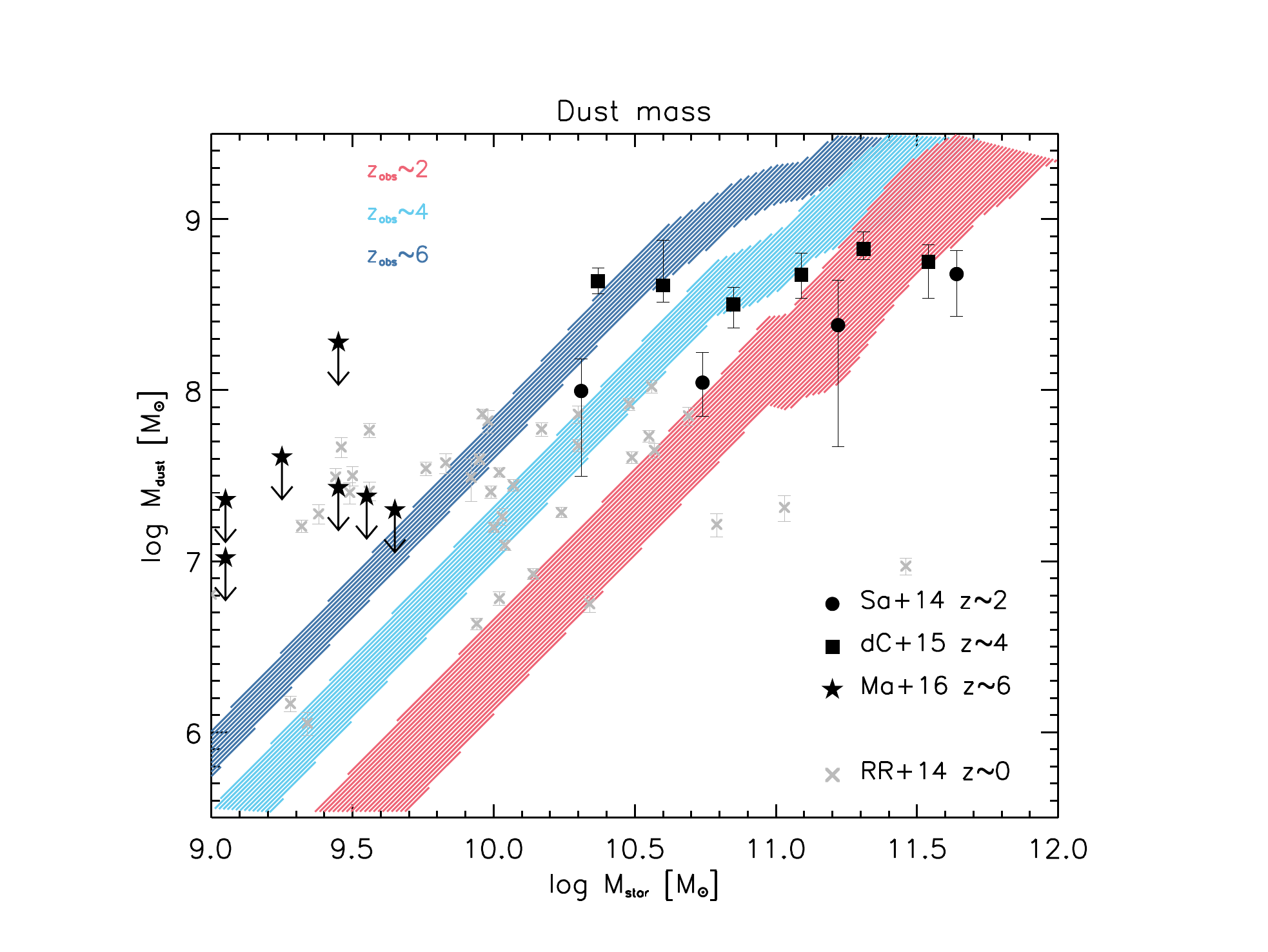}
\caption{Dust mass $M_{\rm dust}$ vs. stellar mass $M_{\star}$, at different observation redshifts $z\approx 2$ (orange), $4$ (cyan), and $6$ (blue); the shaded areas illustrate the $1\sigma$ variance associated to the average over different formation redshifts. Data points are from Santini et al. (2014; circles) at $z\sim 2$, da Cunha et al. (2015; squares) at $z\sim 4$, and from Mancini et al. (2015; stars) $z\sim 6$ (upper limits only); for reference measurements at $z\approx 0$ by Remy-Ruyer et al. (2014; crosses) are also reported.}\label{fig|dustmass}
\end{figure}

In Fig.~\ref{fig|dustmass} we show the dust mass $M_{\rm dust}$ as a function of the stellar mass $M_{\star}$ for different observation redshifts $z_{\rm obs}\sim2$, $4$, and $6$ (color-coded). A direct relationship is expected since both dust and stellar mass are strictly related to the SFR. In fact, the dust mass is produced in SN ejecta/stellar winds, that are more efficient when the SFR is higher; the latter also favors metal rich environments, in turn triggering efficient growth of dust grains by accretion. At given stellar mass, higher redshift galaxies are expected to produce slightly larger amount of dust; the trend can be traced back to their denser environment, in turn yielding shorter star formation timescales.

Our results for starforming ETG progenitors are in good agreement with the observational estimates by Santini et al. (2014) at $z\sim 2$ from stacked far-IR photometry, by da Cunha et al. (2015) at $z\sim 4$ from submm SED modeling, and by Mancini et al. (2015) from upper limits to thermal dust emission in at $z\sim 6$ starforming galaxies.
For reference we also report the $z\approx 0$ dust mass estimates by Remy-Ruyer et al. (2014), mainly referring to disk-dominated galaxies, which are still starforming and moderately gas rich even in the local Universe.

\subsection{Gas metallicity}\label{sec|gasmetal}

\begin{figure}[!ht]
  \centering
\includegraphics[width=1\textwidth]{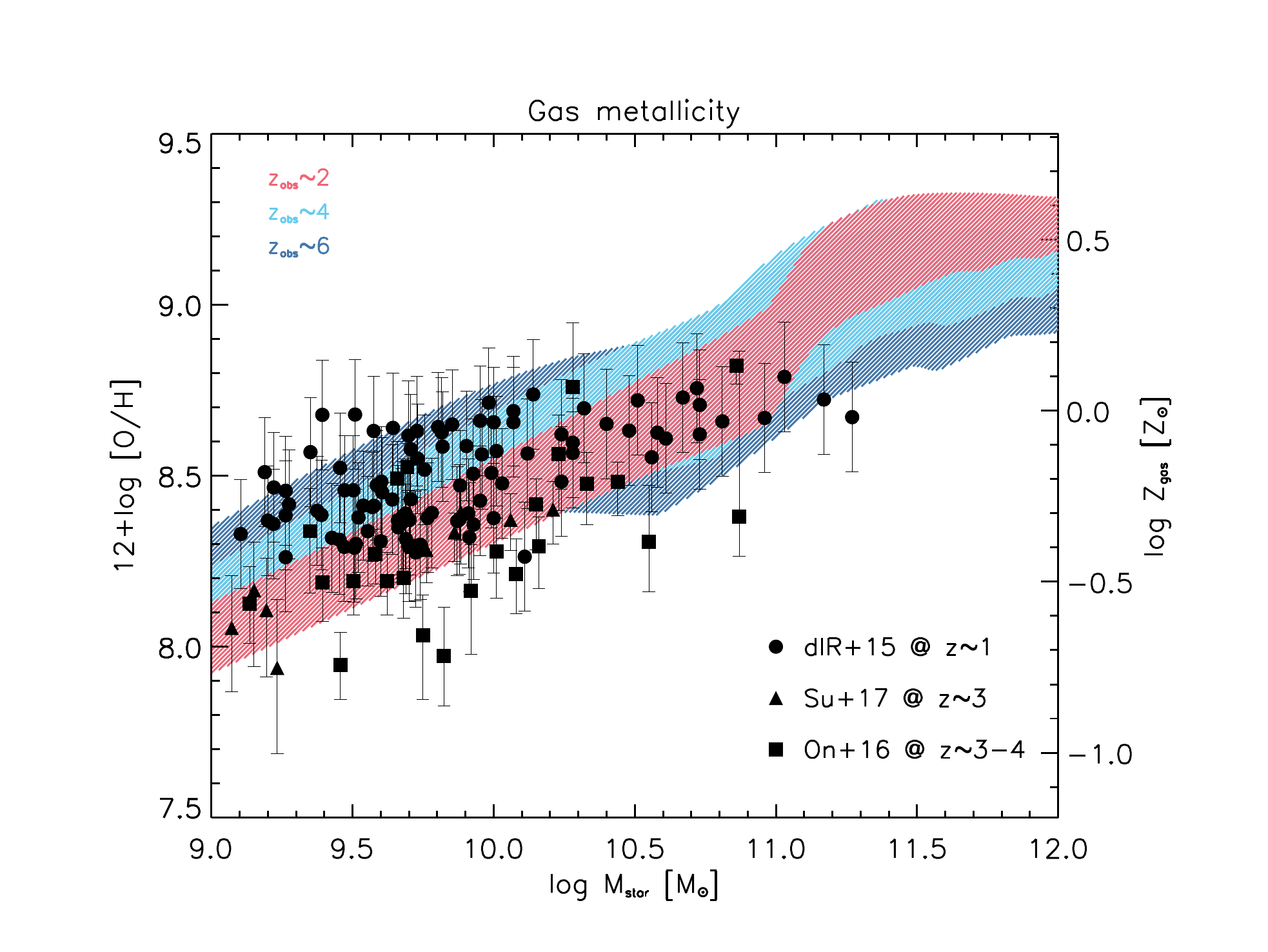}
\caption{Gas metallicity $Z_{\rm gas}$ vs. stellar mass $M_{\star}$ at different observation redshifts $z\approx 2$ (orange), $4$ (cyan), and $6$ (blue); the shaded areas illustrate the $1\sigma$ variance associated to the average over different formation redshifts. Data points are from de los Reyes et al. (2015; circles) at $z\sim1$, Suzuki et al. (2017; triangles) at $z\sim3$ and Onodera et al. (2016; squares) at $z\sim3-4$. All gas metallicity have been converted to PP04O3N2 calibration.}\label{fig|gasmetallicity}
\end{figure}

In Fig.~\ref{fig|gasmetallicity} we present the mass-metallicity relationship (see Tremonti et al. 2004; Erb et al. 2006; Maiolino et al. 2008; Steidel et al. 2014; Zahid et al. 2014; de los Reyes et al. 2015; Sanders et al. 2015; Faisst et al. 2016; Onodera et al. 2016; Suzuki et al. 2017), i.e., the relation between the gas metallicity $Z_{\rm gas}$ of the cold gas and the stellar mass $M_{\star}$ at different observation redshifts $z_{\rm obs}\sim 2$, $4$ and $6$ (color-coded).  The gas metallicity shows an increasing behavior as a function of the final stellar mass, related to the more efficient production of metals in galaxies with higher SFRs, that will also yield larger stellar masses; the corresponding redshift evolution is negligible, being the gas metallicity essentially related to in-situ processes.

Our results are in agreement with gas metallicity estimates (traced mainly by Oxygen abundance, and converted to PP04O3N2 calibration, see Kewley \& Ellison 2008) from strong rest-frame optical emission lines in UV/optically selected starforming galaxies by de los Reyes et al. (2015), Onodera et al. (2016) and Suzuki et al. (2017), spanning the redshift interval $z\sim 1-4$.

\subsection{Stellar metallicity and $\alpha$-enhancement}\label{sec|starmetallicity_sect}

\begin{figure}[!ht]
  \centering
\includegraphics[width=1\textwidth]{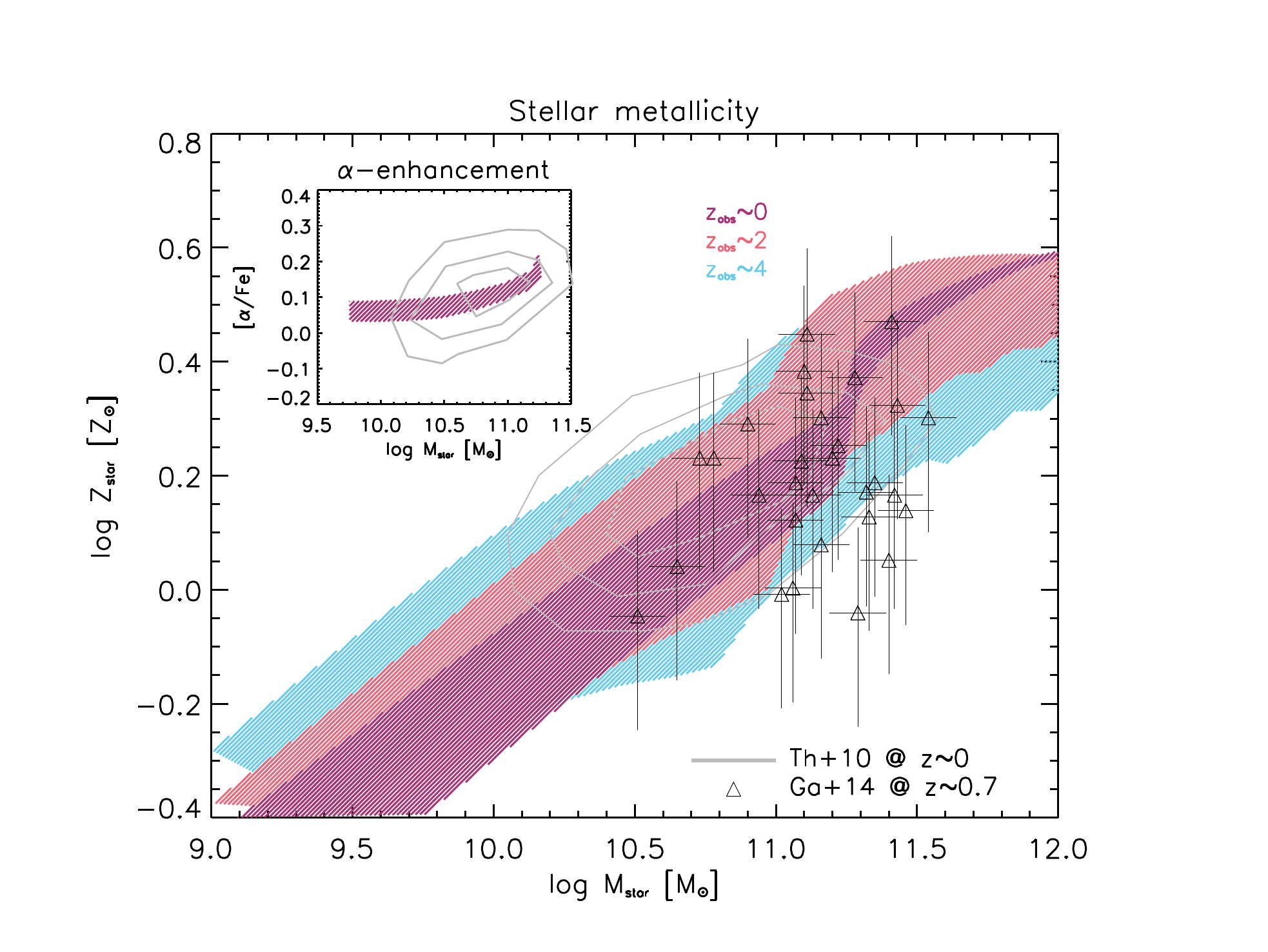}
\caption{Stellar metallicity $Z_{\star}$ vs. stellar mass $M_{\star}$ at different observation redshifts $z\approx 0$ (red), $2$ (orange), and $4$ (cyan); the shaded areas illustrate the $1\sigma$ variance associated to the average over different formation redshifts. Data for SDSS samples of local ETGs are from Thomas et al. (2010; solid contours),  and for individual galaxies at $z\sim 0.7$ are from Gallazzi et al. (2014; triangles). Inset: as above for $\alpha$-elements to iron abundance ratio [$\alpha$/Fe] vs. stellar mass $M_{\star}$ at observation redshift $z\approx 0$ (red).}\label{fig|starmetallicity}
\end{figure}

In Fig.~\ref{fig|starmetallicity} we illustrate the stellar mass-metallicity relationship (see Thomas et al. 2005, 2010; Gallazzi et al. 2006, 2014; Johansson et al. 2012), i.e., the relationship between the stellar metallicity $Z_{\star}$ and the stellar mass, $M_{\star}$ at different observation redshifts $z_{\rm obs}=0$, $2$, and $4$ (color-coded).
The stellar metallicity increases monotonically with stellar mass, mirroring the gas metallicity behaviour (see Fig.~\ref{fig|gasmetallicity}). This is because massive galaxies are characterized on average by higher SFRs, that imply larger stellar masses and metal production. Moreover, in low mass galaxies the depletion of metals by stellar feedback is enhanced due to the shallower potential wells associated to the host halos. Contrariwise, high-mass galaxies can retain greater amounts of chemical-enriched gas, that could be converted and locked into new metal-rich stars, resulting in a higher stellar metallicity. The evolution in redshift is minor, as the stellar metallicity is mainly determined by in-situ star formation processes in the central regions; if any, at higher $z$ and given stellar mass, the stellar metallicity increases slightly since the average SFR is larger.

Our results are in agreement with measurements of stellar metallicity in local ETGs by Thomas et al. (2010) from the SDSS, and with the estimates by Gallazzi et al. (2014) at $z\sim 0.7$, that are broadly consistent with the local relationship within their large uncertainties and intrinsic variance.  Note that other few works attempted to derive stellar metallicity out to $z\sim 3$ in starforming galaxies, but these analysis are based on rest-frame UV absorption features, which are good tracers only of the youngest stellar populations (e.g., Halliday et al. 2008; Sommariva et al. 2012).

In the inset of Fig.~\ref{fig|starmetallicity} we show the $\alpha-$enhancement (see Thomas et al. 2005, 2010; Johansson et al. 2012), i.e., the local $\alpha$-elements to iron abundance ratio [$\alpha$/Fe] as a function of galaxy stellar mass $M_{\star}$. At small stellar masses $M_{\star}\la 10^{10}\,  M_\odot$ an almost constant [$\alpha$/Fe] $\approx 0.05$ is found, while in moving toward higher masses [$\alpha$/Fe] increases up to a value $\sim 0.25$. This trend can be strictly related to the diverse star formation histories characterizing small and high mass galaxies. In particular, in massive galaxies BH feedback is able to quench the star formation and deplete the residual gas mass within a fraction of Gyr, well before type-I$a$ SNe can pollute the ISM  with substantial iron content; this results in an under-abundance of iron with respect to $\alpha$ elements in the stellar component, that in turn cause an excess in [$\alpha$/Fe]. Contrariwise, in low mass galaxies star formation proceeds longer, and type-I$a$ SN have time to enrich the ISM and the stars with relatively larger amount of iron. We find a good agreement of our result with the observed $\alpha$-enhancement of local ETGs, as estimated by Thomas et al. (2010).

\subsection{Outflow metallicity}\label{sec|outmetals}

\begin{figure}[!ht]
  \centering
\includegraphics[width=1\textwidth]{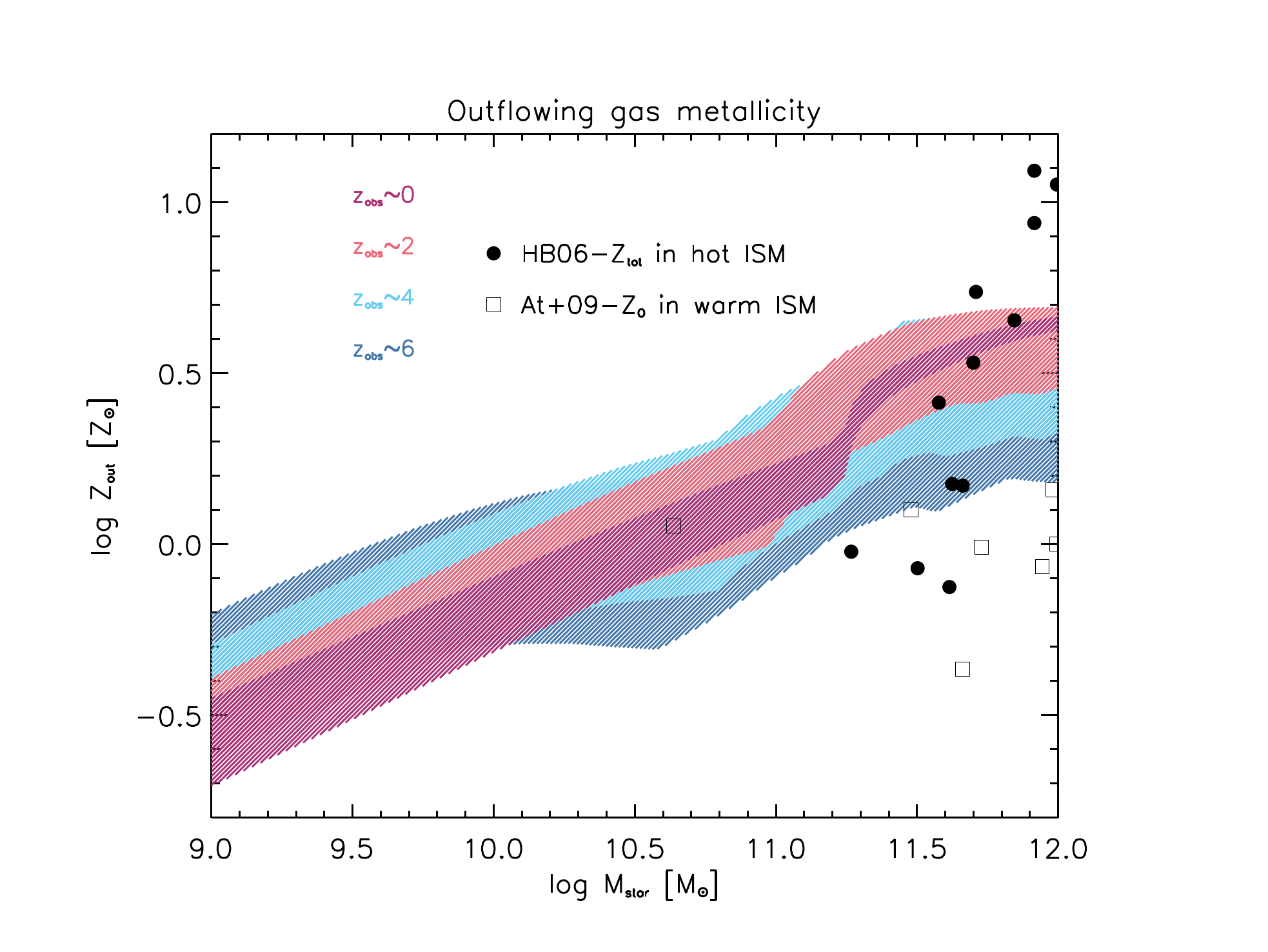}
\caption{Outflowing gas metallicity $Z_{\rm out}$ vs. stellar mass $M_{\star}$ at different observation redshifts $z\approx 0$ (red), $2$ (orange), and $4$ (cyan), and $6$ (blue); the shaded areas illustrate the $1\sigma$ variance associated to the average over different formation redshifts. Data points referring to total $\alpha$-elements abundances in the hot ISM of local ETGs are from from Humphrey \& Buote (2006; filled circles), and
referrring to Oxygen abundance in the warm ISM of local ETGs are from Athey \& Bregman (2009; squares); the latter may be considered lower limits to the total outflowing gas metallicity. Error bars on datapoints of order $\la 1$ dex have been omitted for clarity.}\label{fig|outmetallicity}
\end{figure}

In Fig.~\ref{fig|outmetallicity} we illustrate the metallicity of the outflowing material $Z_{\rm out}$ as a function of the stellar mass $M_{\star}$, for different observation redshifts $z\sim 0$, $2$, $4$, and $6$ (color-coded).
Relevant measurements of the metallicity concern the warm (temperatures $10^4$ K; e.g., Caon et al. 2000; Ferrari et al. 2002; Athey \& Bregman 2009) and hot gas (temperatures $10^6-10^7$ K; see Loewenstein \& Mathews 1991; Mathews \& Brighenti 2003; Humphrey \& Buote 2006) in local massive ETGs, as the outflowing chemical enriched gas has been possibly retained in the host halo potential well. As to the hot gas we report estimates by Humprey \& Buote (2006), who exploited X-ray observations to derive the iron amount and its abundance ratios with other elements in the hot ISM. As to the warm gas we report the estimates by Athey \& Bregman (2009), who determined lower limit to the ISM metallicity of their ETGs sample via oxygen emission lines.

The data points are clustered in the high-mass end of the plot, and are consistent with our results at $z_{\rm obs}\approx 0$, albeit their large uncertainties (omitted for clarity but amounting to about $0.5$ dex) does not allow to draw strong conclusion. Finally, it is worth noticing that the solar or even supersolar metallicities found in the hot and warm medium of local ETGs indicate the ISM metals to have mostly an in-situ, internal origin. As also stressed by Humphrey \& Buote (2006) and Athey \& Bregman (2009) a significant external contribution from cosmic scale primordial gas or minor mergers should imply a significant dilution of the ISM in ETGs, with a ensuing reduction in metallicity. We caveat, however, that the issue is still debated, given the still large observational uncertainties in the ISM metallicity determinations.

\section{Summary}\label{sec|summary}

We have presented a set of new analytic solutions aimed at describing the spatially-averaged evolution of the gas/stellar/dust mass and metal content in a starforming galaxy hosted within a dark halo of given mass and formation redshift (see Sect.~\ref{sec|An_sol} and Figs.~\ref{fig|timevo_z3} and \ref{fig|timevo_z6}). The basic framework pictures the galaxy as an open, one-zone system comprising three interlinked mass components: a reservoir of warm gas subject to cooling and condensation toward the central regions; cold gas fed by infall and depleted by star formation and stellar feedback (type-II SNe and stellar winds); stellar mass, partially restituted to the cold phase by stars during their evolution. The corresponding metal enrichment history of the cold gas and stellar mass is self-consistently computed using as input the solutions for the evolution of the mass components; the metal equations includes effects of feedback, astration, instantaneous production during star formation, and delayed production by type-I$a$ SNe, possibly following a specified delay time distribution. Finally, the dust mass evolution takes into account the formation of grain cores associated to star formation, and of the grain mantles due to accretion onto pre-existing cores; astration of dust by star formation and stellar feedback, and spallation by SN shockwaves are also included.

We have then applied our analytic solutions to describe the formation of ETGs and the evolution of their starforming progenitors (see Sect.~\ref{sec|Application}). To this purpose, we have supplemented our solutions with a couple of additional ingredients: (i) specific prescriptions for parameter setting, inspired by in-situ galaxy-black hole coevolution scenarios for ETG formation; (ii) estimates of the average halo and stellar mass growth by mergers, computed on the basis of the merger rates from state-of-the-art numerical simulations (see Fig.~\ref{fig|merging}). We then derive a bunch of fundamental relationships involving spatially-averaged quantities as a function of the observed stellar mass: star formation efficiency (see Fig.~\ref{fig|fstar}), SFR (see Fig.~\ref{fig|MS}), gas mass (see Fig.~\ref{fig|gasmass}), dust mass (see Fig.~\ref{fig|dustmass}), gas metallicity (see Fig.~\ref{fig|gasmetallicity}), stellar metallicity and $[\alpha$/Fe$]$ ratio (see Fig.~\ref{fig|starmetallicity}), and outflowing gas metallicity (see Fig.~\ref{fig|outmetallicity}). We compare these relationships with the data concerning local ETGs and their high-$z$  starforming progenitors, finding a pleasing overall agreement (see Sect.~\ref{sec|Results}). We remark that a major value of our approach is to reproduce, with a unique set of physically motivated parameters, a wealth of observables concerning ETGs and their progenitors.

Another straightforward application of our solutions would be the description of the spatially-averaged properties for local disk-dominated (e.g., spiral) galaxies. On the one hand, this will just require different prescriptions for parameter setting with respect to ETG progenitors. In a nutshell, we expect appreciably longer condensation/starformation timescales of the order of several Gyrs, and a minor role of BH feedback; this will originate a prolonged star formation history to low SFR levels and appreciable dilution from infalling matter, in turn implying slower accumulation of stellar mass, metals, and dust. On the other hand, additional processes like galactic fountains, differential winds, stellar mixing and multi-zonal effects may play a relevant role in local spirals; this will require the basic framework presented in Sect.~\ref{sec|An_sol} to become more complex, and the search for realistic analytic solutions more challenging. We will pursue this program in a forthcoming paper.

To sum up, the analytic solutions provided here are based on an idealized albeit non-trivial descriptions of the main physical processes regulating galaxy formation on a spatially-average ground, that go beyond simple approaches to the history of star formation and chemical enrichment like the closed/leaky box or the gas regulator model (see Appendix A and Fig.~\ref{fig|timevo_comp}). Yet, our solutions are simple enough to easily disentangle the role of the main physical processes at work, to allow a quick exploration of the parameter space, and to make transparent predictions on spatially-averaged quantities. All in all, our analytic solutions may provide a basis for improving the (subgrid) physical recipes implemented in theoretical approaches and numerical simulations, and can offer a benchmark for interpreting and forecasting current and future observations with multiband coverage, that will become routinely achievable even at high redshift, e.g., via targeted observations with ALMA and via dedicated surveys with the JWST.

\begin{appendix}

\section{Comparison with classic analytic models}

In this Appendix we compare our new analytic solutions with classic analytic models, extensively adopted to describe the spatially-averaged behavior of starforming galaxies. We focus on two classes of models, that have as limiting cases the classical closed-box and gas regulator (or bathtube) solutions.

\subsection{No inflow models and the leaky/closed-box solution}

A class of simple analytic models can be obtained by imposing no inflow of gas. The relevant equations describing the evolution of gas mass and metallicity can be written as
\begin{equation}
\left\{
\begin{aligned}
\dot M_{\mathrm{cold}} &= - \gamma\, \dot M_{\star}~,\\
\\
\dot Z_{\rm cold} &=\frac{y_Z\,(1-\mathcal{R})}{\gamma}\, \frac{\dot{M}_{\rm cold}}{M_{\rm cold}}~,\\
\end{aligned}
\right.
\end{equation}
with initial conditions $M_{\rm cold}(0)=M_{\rm cold,0}$ and $Z_{\rm cold}(0)=0$. The interesting feature of this class of models is that, irrespective of the shape of the SFR $\dot M_{\star}$ and of its relation with the gas mass $M_{\rm cold}$, an implicit solution can be provided in terms of the gas mass fraction $\mu\equiv M_{\rm cold}/(M_{\rm cold}+M_\star)$. Solving the above equations in terms of $\mu$ one easily obtains
\begin{equation}\label{eq|noinflow}
\left\{
\begin{aligned}
M_{\mathrm{cold}} &= M_{\rm cold,0}\, \frac{\mu}{ \gamma-(\gamma-1)\, \mu}~,\\
\\
Z_{\rm cold} &=\frac{y_Z\,(1-\mathcal{R})}{\gamma}\, \ln\left(\frac{1}{\mu}-\frac{\gamma-1}{\gamma}\right)~.\\
\end{aligned}
\right.
\end{equation}
At late times one expects $\mu\ll 1$, hence the metallicity asymptotically behaves as $Z_{\rm cold}\simeq y_Z\, (1-\mathcal{R})\, \gamma^{-1}\ln \mu^{-1}$; this is often referred to as the leaky-box solution.
For $\gamma\simeq 1$ (i.e., $\epsilon_{\rm out}=\mathcal{R}=0$), which correspond to neglecting outflow and recycling, one finds the classical closed-box solution; in such a case the total mass is constant in time and hence $\mu\simeq M_{\mathrm{cold}}/M_{\rm cold,0}$ and $Z_{\rm cold}\simeq y_Z\,\ln \mu^{-1}$.

Coming back to the solution in Eqs.~(\ref{eq|noinflow}), one can find an explicit time dependence by specifying the relation between the SFR and the gas mass; in the way of comparing the outcome with our result of Eqs.~(\ref{eq|basicsol}) and (\ref{eq|Zcold}) in the main text, we prescribe as in Eqs.~(\ref{eq|basics}) that $\dot M_\star=M_{\rm cold}/\tau_{\star}=s\, M_{\rm cold}/\tau_{\rm cond}$, and that $M_{\rm cold,0}=f_{\rm inf}\, M_b$. In such a case it is found that
\begin{equation}\label{eq|noinflowsol}
\left\{
\begin{aligned}
M_{\mathrm{cold}}(x) &= f_{\rm inf}\, M_b\, e^{-s\gamma\, x}~,\\
\\
Z_{\rm cold}(x) &=\frac{y_Z\,(1-\mathcal{R})}{\gamma}\, \left(s\gamma\,x-\ln{\gamma}\right)~,\\
\end{aligned}
\right.
\end{equation}
with $x\equiv \tau/\tau_{\rm cond}$. Clearly, the absence of dilution due to inflow makes the metallicity to increase almost linearly with galactic age, while the cold gas reservoir diminishes and gets progressively exhausted. In particular, for the closed-box model ($\gamma\simeq 1$), one has $M_{\rm cold}\propto e^{-s\,\tau/\tau_{\rm cond}}$ and $Z_{\rm cold}\propto y_Z\,\tau/\tau_{\rm cond}$. The evolution of the gas mass and metallicity is illustrated in Fig.~\ref{fig|timevo_comp}.

\subsection{Constant inflow models and the gas-regulator (or bathtube) solution}

Another class of more realistic analytic models can be obtained when the inflow of gas is assumed to occur at a constant rate $\dot M_{\rm inf}$. The relevant evolution equations reads
\begin{equation}\label{eq|costinflow}
\left\{
\begin{aligned}
\dot M_{\mathrm{cold}} &= \dot M_{\rm inf}- \gamma\, \dot M_{\star}~,\\
\\
\dot Z_{\rm cold} &=-\frac{\dot M_{\rm inf}}{M_{\rm cold}}\, Z_{\rm cold}+y_Z\,(1-\mathcal{R})\, \frac{\dot M_\star}{M_{\rm cold}}~,\\
\end{aligned}
\right.
\end{equation}
with initial conditions $M_{\rm cold}(0)=Z_{\rm cold}(0)=0$. To solve the system a prescription linking the SFR to the gas mass is needed; in the way of comparing with our result we adopt $\dot M_\star=M_{\rm cold}/\tau_\star=s\, M_{\rm cold}/\tau_{\rm cond}$ as in Eqs.~(\ref{eq|basics}).

The remarkable aspect of the constant inflow model is that it admits self-consistent steady state solutions for both the cold gas mass and the cold gas metallicity. In steady state, the SFR adjusts such that the mass loss due to feedback and the mass addition due to inflow exactly compensate, and at the same time the metal dilution due to inflow and the metal production due to star formation also balance. Posing $\dot M_{\rm cold}=\dot Z_{\rm cold}=0$ in the equations above one finds
\begin{equation}\label{eq|bathtube}
\left\{
\begin{aligned}
\bar M_{\rm cold} & \simeq \frac{\dot M_{\rm inf}\, \tau_{\rm cond}}{s \gamma}~,\\
\\
\bar Z_{\rm cold} & \simeq \frac{y_Z\,(1-\mathcal{R})}{\gamma}~.\\
\end{aligned}
\right.
\end{equation}
This steady-state solution is known as gas-regulator or bathtube model. In the context of galaxy formation, the inflow rate $\dot M_{\rm inf}$ is assumed to originate via continuous accretion of gas from the environment outside the host halo; this is reasonable for galaxies which have long star formation timescales like local spirals, while clearly it cannot be retained for high-redshift starforming galaxies and specifically for ETG progenitors.

\begin{figure}[!ht]
  \centering
\includegraphics[width=1\textwidth]{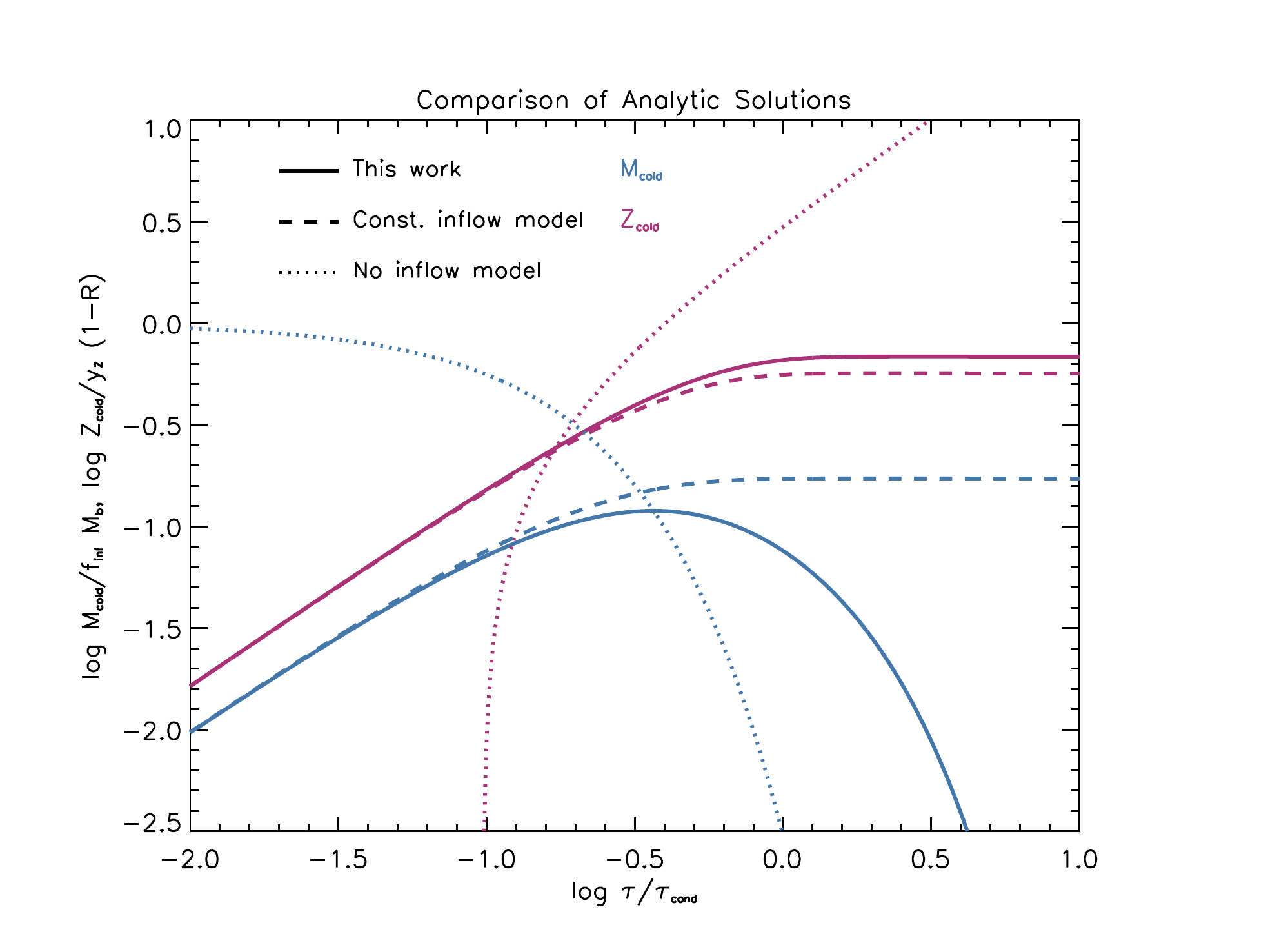}
\caption{Comparison of the analytic solutions developed in this work (solid lines) with the classical no inflow/closed/leaky-box model (dotted lines) and constant inflow/gas regulator model (dashed lines), see Appendix A for more details. Blue lines illustrate the time evolution of the cold gas mass $M_{\rm cold}/f_{\rm inf}\,M_b$, and red lines that of the cold gas metallicity $Z_{\rm cold}/y_Z\,(1-\mathcal{R})$. For a sensible comparison, we have adopted values of the relevant parameters $s\approx 3$ and $\gamma\approx 2$ apt for a galaxy hosted in a halo with mass $M_{\rm H}\approx 10^{12}\, M_\odot$ formed at $z_{\rm form}\approx 3$.}\label{fig|timevo_comp}
\end{figure}

The steady state solution is an attractor, i.e., in the long run the system converges to it; the overall evolution, derived by solving the Eqs.~(\ref{eq|costinflow}), is given by
\begin{equation}\label{eq|gasinflowsol}
\left\{
\begin{aligned}
M_{\rm cold}(x) & = \frac{f_{\rm inf}\, M_{\rm b}}{s \gamma}\, (1-e^{-s\gamma\,x})~,\\
\\
Z_{\rm cold}(x) & = \frac{y_Z\,(1-\mathcal{R})}{\gamma}\, \left[1-\frac{s\gamma\, x}{e^{s\gamma\, x}-1}\right]~,\\
\end{aligned}
\right.
\end{equation}
where $x\equiv \tau/\tau_{\rm cond}$, and for easing the comparison with our solutions in Eqs.~(\ref{eq|basicsol}) and (\ref{eq|Zcold}) we have prescribed $\dot M_{\rm inf}=f_{\rm inf}\, M_{\rm b}/\tau_{\rm cond}$. The evolution of the gas mass and metallicity is illustrated in Fig.~\ref{fig|timevo_comp}.

\end{appendix}

\begin{acknowledgements}
We acknowledge the anonymous referee for a constructive report. We thank A. Bressan, A. Feltre, D. Donevski, and G. Rodighiero for helpful discussions. This work has been partially supported by PRIN MIUR 2017 prot.
20173ML3WW 002 Opening the ALMA window on the cosmic evolution of gas, stars and supermassive black holes. AL acknowledges the MIUR grant 'Finanziamento annuale individuale attivit\'a base di ricerca'.
\end{acknowledgements}

\clearpage

\clearpage

\clearpage

\clearpage

\clearpage

\clearpage

\clearpage

\clearpage

\clearpage

\clearpage

\clearpage

\begin{references}
\reference{}Allende Prieto, C., Lambert, D. L., \& Asplund, M. 2001, ApJ, 556, L63
\reference{}Andrews, B.H., Weinberg, D.H., Schonrich, R., Johnson, J.A. 2017, ApJ, 835, 224
\reference{}Aoyama, S., Hou, K.-C., Hirashita, H, Nagamine, K., \& Shimizu, I. 2018, MNRAS, 478, 4905
\reference{}Aoyama, S., Hou, K.-C., Shimizu, I. 2017, MNRAS, 466, 105
\reference{}Arrigoni, M., Trager, S.C., Somerville, R.S., \& Gibson, B.K. 2010, MNRAS, 402, 173
\reference{}Asano, R.S., Takeuchi, T.T., Hirashita, H., Inoue, A.K. 2013, EP\&S, 65, 213
\reference{}Athey, A.E., \& Bregman, J.N. 2009, ApJ, 696, 681
\reference{}Aversa, R., Lapi, A., De Zotti, G., Shankar, F., \& Danese, L. 2015, ApJ, 810, 74
\reference{}Barnes, J., \& Efstathiou, G. 1987, ApJ, 319, 575
\reference{}Barro, G., Kriek, M., Pérez-González, P. G., et al. 2016, ApJL, 827, L32
\reference{}Barro, G., Kriek, M., Perez-Gonzalez, P. G., et al. 2017, ApJL, 851, L40
\reference{}Behroozi, P.S., Wechsler, R.H., \& Conroy, C. 2013, ApJ, 770, 57
\reference{}Bekki, K. 2015, MNRAS, 449, 1625
\reference{}Bekki, K. 2013, MNRAS, 436, 2254
\reference{}Bell, E. F., McIntosh, D. H., Katz, N., \& Weinberg, M. D. 2003, ApJS, 149, 289
\reference{}Benson, A.J. 2012, NewA, 17, 175
\reference{}Bianchi, S., \& Schneider, R. 2007, MNRAS, 378, 973
\reference{}Bigiel, F., Leroy, A., Walter, F., et al. 2008, AJ, 136, 2846
\reference{}Bocquet, S., Saro, A., Dolag, K., \& Mohr, J.J. 2016, MNRAS, 456, 2361
\reference{}Boquien, M., Burgarella, D., Roehlly, Y., et al. 2019, A\&A, 622, A103
\reference{}Bouche, N., Dekel, A., Genzel, R., et al. 2010, ApJ, 718, 1001
\reference{}Bournaud, F. 2016, Galactic Bulges (Switzerland: Springer International Publishing), p. 355
\reference{}Bregman, J. N. 1980, ApJ, 365, 544
\reference{}Bullock, J. S., Dekel, A., Kolatt, T. S., et al. 2001, ApJ, 555, 240
\reference{}Burkert, A., Forster Schreiber, N. M., Genzel, R., et al. 2016, ApJ, 826, 214
\reference{}Caon, N., Macchetto, D., \& Pastoriza, M. 2000, ApJS, 127, 39
\reference{}Cassar\'a, L. P., Maccagni, D., Garilli, B., et al. 2016, A\&A, 593, A9
\reference{}Chabrier, G. 2005, in The Initial Mass Function 50 years later, Astrop. Sp. Sci. 327, ed. by E. Corbelli and F. Palle (Dordrecht: Springer), p. 41
\reference{}Chabrier, G. 2003, ApJL, 586, L133
\reference{}Chiappini, C., Matteucci, F., \& Romano, D. 2001, ApJ, 554, 1044
\reference{}Chiosi, C. 1980, A\&A, 83, 206
\reference{}Citro, A., Pozzetti, L., Moresco, M., \& Cimatti, A. 2016, A\&A, 592, A19
\reference{}Cole, S., Lacey, C.G., Baugh, C.M., \& Frenk, C.S. 2000, MNRAS, 319, 168
\reference{}Collacchioni, F., Cora, S.A., Lagos, C.D.P., \& Vega-Martinez, C.A. 2018, MNRAS, 481, 954
\reference{}Comparat, J., Prada, F., Yepes, G., \& Klypin, A. 2019, MNRAS, 483, 2561
\reference{}Comparat, J., Prada, F., Yepes, G., \& Klypin, A. 2017, MNRAS, 469, 4157
\reference{}Conroy, C. 2013, ARA\&A, 51, 393
\reference{}Cousin, M., Buat, V., Boissier, S., et al. 2016, A\&A, 589, A109
\reference{}Croton, D.J., Springel, V., White, S.D.M., et al. 2006, MNRAS, 365, 11
\reference{}da Cunha, E., Walter, F., Smail, I.R., et al. 2015, ApJ, 806, 110
\reference{}Daddi, E., Dickinson, M., Morrison, G., et al. 2007, ApJ, 670, 156
\reference{}Danovich, M., Dekel, A., Hahn, O., Ceverino, D., \& Primack, J. 2015, MNRAS, 449, 2087
\reference{}Dav\'e, R., Finlator, K., \& Oppenheimer, B.D. 2012, MNRAS, 421, 98
\reference{}de Bennassuti M., Schneider R., Valiante R., \& Salvadori S., 2014, MNRAS, 445, 3039
\reference{}Dekel, A., Lapiner, S., \& Dubois, Y. 2019, A\&A, submitted [arXiv:1904.08431]
\reference{}Dekel, A., \& Mandelker, N. 2014, MNRAS, 444, 2071
\reference{}Dekel, A., Birnboim, Y., Engel, G., et al. 2009, Natur, 457, 451
\reference{}de los Reyes, M.A., Ly, C., Lee, J.C., et al. 2015, AJ, 149, 79
\reference{}De Lucia, G., Fontanot, F., \& Hirschmann, M. 2017, MNRAS, 466, L88
\reference{}De Lucia, G., Tornatore, L., Frenk, C.S., et al. 2014, MNRAS, 445, 970
\reference{}De Rossi, M.E., \& Bromm, V. 2019, ApJ, submitted [arXiv:1903.02512]
\reference{}Draine, B. T. 2003, ARA\&A, 41, 241
\reference{}Draine, B. T. 2011, Physics of the Interstellar and Intergalactic Medium
(Princeton: Princeton Univ. Press; NJ, USA)
\reference{}Dubois, Y., Peirani, S., Pichon, C. 2016, MNRAS, 463, 3948
\reference{}Dubois, Y., Pichon, C., Welker, C., et al. 2014, MNRAS, 444, 1453
\reference{}Dunlop, J. S., McLure, R. J., Biggs, A. D., et al. 2017, MNRAS, 466, 861
\reference{}Dwek, E., \& Cherchneff, I. 2011, ApJ, 727, 63
\reference{}Dwek, E. 1998, ApJ, 501, 643
\reference{}Dye, S., Furlanetto, C., Swinbank, A. M., et al. 2015, MNRAS, 452, 2258
\reference{}Edmunds, M. G. 1990, MNRAS, 246, 678
\reference{}Elbaz, D., Daddi, E., Le Borgne, D., et al. 2007, A\&A, 468, 33
\reference{}Elmegreen, D. M., Elmegreen, B. G., \& Ferguson, T. E. 2005, ApJL, 623, L71
\reference{}Erb, D.K. 2008, ApJ, 674, 151
\reference{}Erb, D. K., Shapley, A. E., Pettini, M., et al. 2006, ApJ, 644, 813
\reference{}Fabian, A. C. 1999, MNRAS, 308, L39
\reference{}Faisst, A. L., Capak, P. L., Davidzon, I., et al. 2016, ApJ, 822, 29
\reference{}Fakhouri, O., Ma, C.-P., \& Boylan-Kolchin, M. 2010, MNRAS, 406, 2267
\reference{}Fakhouri, O., \& Ma, C.-P. 2008, MNRAS, 386, 577
\reference{}Feldmann, R. 2015, MNRAS, 449, 3274
\reference{}Ferrari, F., Pastoriza, M. G., Macchetto, F. D., et al. 2002, A\&A, 389, 355
\reference{}Finlator, K., Oh, S. P., Ozel, F., \& Dave, R. 2012, MNRAS, 427, 2464
\reference{}Fontanot, F., De Lucia, G., Hirschmann, M., et al. 2017, MNRAS, 464, 3812
\reference{}Forbes, J. C., Krumholz, M. R., \& Speagle, J. S. 2019, MNRAS, 487, 3581
\reference{}Forbes, J.C., Krumholz, M.R., Burkert, A., \& Dekel, A. 2014a, MNRAS, 443, 168
\reference{}Forbes, J.C., Krumholz, M.R., Burkert, A., \& Dekel, A. 2014b, MNRAS, 438, 1552
\reference{}Fu, J., Kauffmann, G., Huang, M.-l., et al. 2013, MNRAS, 434, 1531
\reference{}Gallazzi, A., Bell, E. F., Zibetti, S., Brinchmann, J., \& Kelson, D. D. 2014, ApJ, 788, 72
\reference{}Gallazzi, A., Charlot, S., Brinchmann, J., \& White, S. D. M. 2006, MNRAS, 370, 1106
\reference{}Genel, S., Vogelsberger, M., Springel, Volker, et al. 2014, MNRAS, 445, 175
\reference{}Genel, S., Bouch\'e, N., Naab, T., Sternberg, A., \& Genzel, R. 2010, ApJ, 719, 229
\reference{}Genzel, R., Newman, S., Jones, T., et al. 2011, ApJ, 733, 101
\reference{}Giocoli, C., Tormen, G., \& Sheth, R.K. 2012, MNRAS, 422, 185
\reference{}Gioannini, L., Matteucci, F., Vladilo, G., \& Calura, F. 2017, MNRAS, 464, 985
\reference{}Granato, G. L., De Zotti, G., Silva, L., Bressan, A., \& Danese, L. 2004, ApJ, 600, 580
\reference{}Greggio, L. 2005, A\&A, 441, 1055
\reference{}Grisoni, V., Spitoni, E., \& Matteucci, F. 2018, MNRAS, 481, 2570
\reference{}Halliday, C., Daddi, E., Cimatti, A., et al. 2008, A\&A, 479, 417
\reference{}Hartwick, F.D.A. 1976, ApJ, 209, 418
\reference{}Hirashita, H., Nozawa, T., Villaume, A., \& Srinivasan, S. 2015, MNRAS, 454, 1620
\reference{}Hirashita, H. 2000, PASJ, 52, 585
\reference{}Hirschmann, M., De Lucia, G., \& Fontanot, F. 2016, MNRAS, 461, 1760
\reference{}Hodge, J. A., Swinbank, A. M., Simpson, J. M., et al. 2016, ApJ, 833, 103
\reference{}Hopkins, P.F., Wetzel, A., Keres, D., et al. 2018, MNRAS, 480, 800
\reference{}Hopkins, P.F., Keres, D., Onorbe, J., et al. 2014, MNRAS, 445, 581
\reference{}Hopkins, P. F., Quataert, E., \& Murray, N. 2012, MNRAS, 421, 3522
\reference{}Hudson, M. J., Gillis, B. R., Coupon, J., et al. 2015, MNRAS, 447, 298
\reference{}Humphrey, P.J., \& Buote, D.A. 2006, ApJ, 639, 136
\reference{}Ikarashi, S., Ivison, R. J., Caputi, K. I., et al. 2015, ApJ, 810, 133
\reference{}Iliev, I. T., Mellema, G., Shapiro, P. R., \& Pen, U.-L. 2007, MNRAS, 376, 534
\reference{}Imara, N., Loeb, A., Johnson, B.D., Conroy, C., \& Behroozi, P. 2018, ApJ, 854, 36
\reference{}Inoue, A.K., 2003, PASJ, 55, 901
\reference{}Jiang, F., Dekel, A., Kneller, O., et al. 2019, MNRAS, submitted [arXiv:1804.07306]
\reference{}Johansson, J., Thomas, D., \& Maraston, C. 2012, MNRAS, 421, 1908
\reference{}Kauffmann, G., White, S. D. M., \& Guiderdoni, B. 1993, MNRAS, 264, 201
\reference{}Kennicutt, R. C., Jr. 1998, ApJ, 498, 541
\reference{}Kewley, L.J., \& Ellison, S.L. 2008, ApJ, 681, 1183
\reference{}King, A. R. 2014, SSRv, 183, 427
\reference{}King, A. R. 2003, ApJL, 596, L27
\reference{}Kitayama, T., \& Suto, Y. 1996, MNRAS, 280, 638
\reference{}Koprowski, M., Dunlop, J. S., Michalowski, M. J., et al. 2016, MNRAS, 458, 4321
\reference{}Kravtsov A., Vikhlinin A., \& Meshscheryakov A. 2014 [arXiv:1401.7329]
\reference{}Krumholz, M. R., Dekel, A., \& McKee, C. F. 2012, ApJ, 745, 69
\reference{}Kubryk, M., Prantzos, N., \& Athanassoula, E. 2015, A\&A, 580, A126
\reference{}Lacey, C.G., Baugh, C.M., Frenk, C.S., et al. 2016, MNRAS, 462, 3854
\reference{}Lacey, C., \& Cole, S. 1993, MNRAS, 262, 627
\reference{}Lacey, C. G., \& Fall, M. 1985, ApJ, 290, 154
\reference{}Lada, C.J., Lombardi, M., \& Alves, J.F. 2010, ApJ, 724, 687
\reference{}Lagos, C.d.P., Tobar, R.J., Robotham, A.S.G., et al. 2018, MNRAS, 481, 3573
\reference{}Lang, P., Schinnerer, E., Smail, I., et al. 2019, ApJ, 879, 54
\reference{}Lapi, A., Pantoni, L., Zanisi, L., et al. 2018, ApJ, 857, 22
\reference{}Lapi, A., Mancuso, C., Bressan, A., \& Danese, L. 2017, ApJ, 847, 13
\reference{}Lapi, A., Raimundo, S., Aversa, R., et al. 2014, ApJ, 782, 69
\reference{}Lapi, A., Salucci, P., \& Danese, L. 2013, ApJ, 772, 85
\reference{}Lapi, A., Shankar, F., Mao, J., et al. 2006, ApJ, 650, 42
\reference{}Lara-Lopez, M.A., Lopez-Sanchez, Á.R., \& Hopkins, A.M. 2013, ApJ, 764, 178
\reference{}Lilly, S.J., Carollo, C. M., Pipino, A., Renzini, A., \& Peng, Y. 2013, ApJ, 772, 119
\reference{}Loewenstein, M., \& Mathews, W.G. 1991, ApJ, 373, 445
\reference{}Ma, J., Gonzalez, A. H., Viera, J. D., et al. 2016, ApJ, 832, 114
\reference{}Mac Low, M.-M., \& Ferrara, A. 1999, ApJ, 513, 142
\reference{}Macci\'o, A. V., Dutton, A. A., van den Bosch, F. C., et al. 2007, MNRAS, 378, 55
\reference{}Maiolino, R., \& Mannucci, F. 2019, A\&ARv, 27, 3
\reference{}Maiolino, R., Nagao, T., Grazian, A., et al. 2008, A\&A, 488, 463
\reference{}Maiolino, R., Cox, P., Caselli, P., et al. 2005, A\&A, 440, L51
\reference{}Man, A. W. S., Greve, T. R., Toft, S., et al. 2016, ApJ, 820, 11
\reference{}Mancini, M., Schneider, R., Graziani, L., et al. 2015, MNRAS, 451, L70
\reference{}Mancuso, C., Lapi, A., Shi, J., et al. 2016a, ApJ, 823, 128
\reference{}Mancuso, C., Lapi, A., Shi, J., et al. 2016b, ApJ, 833, 152
\reference{}Mandelbaum, R., Wang, W., Zu, Y., et al. 2016, MNRAS, 457, 3200
\reference{}Mannucci, F., Cresci, G., Maiolino, R., Marconi, A., \& Gnerucci, A. 2010, MNRAS, 408, 2115
\reference{}Mannucci, F., Della Valle, M, \& Panagia, N. 2006, MNRAS, 370, 773
\reference{}Maoz, D., \& Graur, O. 2017, ApJ, 848, 25
\reference{}Maoz D., Mannucci F., \& Nelemans G. 2014, ARA\&A, 52, 107
\reference{}Marconi, G., Matteucci, F., \& Tosi, M. 1994, MNRAS, 270, 35
\reference{}Mathews, W.G., \& Brighenti, F. 2003, ARA\&A, 41, 191
\reference{}Matteucci, F. 2012, Chemical Evolution of Galaxies (Berlin Heidelberg: Springer-Verlag)
\reference{}Matteucci, F., \& Greggio, L. 1986, A\&A, 154, 279
\reference{}Matteucci, F., \& Chiosi, C. 1983, A\&A, 123, 121
\reference{}McAlpine, S., Smail, I., Bower, R.G., et al. 2019, MNRAS, submitted [arXiv:1901.05467]
\reference{}McAlpine, S., Helly, J.C., Schaller, M., et al. 2016, A\&C, 15, 72
\reference{}McKee C., 1989, in Allamandola L. J., Tielens A. G. G. M., eds, Proc. IAU Symp. 135, Interstellar Dust. Kluwer, Dordrecht, p. 431
\reference{}McKinnon, R., Vogelsberger, M., Torrey, P., Marinacci, F., \& Kannan, R.2018, MNRAS, 478, 2851
\reference{}McKinnon, R., Torrey, P., Vogelsberger, M., Hayward, C.C., \& Marinacci, F. 2017, MNRAS, 468, 1505
\reference{}Michalowski, M.J. 2015, A\&A, 577, A80
\reference{}Mo, H., van den Bosch, F., \& White, S.D.M. 2010, Galaxy Formation and Evolution (Cambridge: Cambridge Univ. Press)
\reference{}Mo, H. J., \& Mao, S. 2004, MNRAS, 353, 829
\reference{}Moll\'a, M., Cavichia, O., Gavilan, M., \& Gibson, B.K. 2015, MNRAS, 451, 3693
\reference{}More, S., van den Bosch, F. C., Cacciato, M., et al. 2011, MNRAS, 410, 210
\reference{}Moreno, J., Giocoli, C., \& Sheth, R. K. 2009, MNRAS, 397, 299
\reference{}Moster, B. P., Naab, T., \& White, S. D. M. 2013, MNRAS, 428, 3121
\reference{}Moustakas, J., Coil, A. L., Aird, J., et al. 2013, ApJ, 767, 50
\reference{}Murray, N., Quataert, E., \& Thompson, T. A. 2005, ApJ, 618, 569
\reference{}Naab, T., \& Ostriker, J.P. 2017, ARA\&A, 55, 59
\reference{}Naab, T., \& Ostriker, J.P. 2006, MNRAS, 366, 899
\reference{}Navarro, J. F., Frenk, C. S., \& White, S. D. M. 1997, ApJ, 490, 493
\reference{}Negrello, M., Hopwood, R., Dye, S., et al. 2014, MNRAS, 440, 1999
\reference{}Noeske, K. G., Weiner, B. J., Faber, S. M., et al. 2007, ApJ, 660, L43
\reference{}Nomoto, K., Kobayashi, C., \& Tominaga, N. 2013, ARA\&A, 51, 457
\reference{}Omont, A., Willott, C. J., Beelen, A., et al. 2013, A\&A, 552, A43
\reference{}Onodera, M., Carollo, C. M., Lilly, S. et al. 2016, ApJ, 822, 42
\reference{}Oppenheimer, B.D., \& Dave, R. 2006, MNRAS, 373, 1265
\reference{}Pagel, B. E. J., \& Patchett, B. E. 1975, MNRAS, 172, 13
\reference{}Pallottini, A., Ferrara, A., Gallerani, S., et al. 2017, MNRAS, 465, 2540
\reference{}Papovich, C., Finkelstein, S. L., Ferguson, H. C., Lotz, J. M., \& Giavalisco, M. 2011, MNRAS, 412, 1123
\reference{}Pawlik, A. H., Schaye, J., \& van Scherpenzeel, E. 2009, MNRAS, 394, 1812
\reference{}Pezzulli, G., \& Fraternali, F. 2016, AN, 337, 913
\reference{}Pilyugin, L. S. 1993, A\&A, 277, 42
\reference{}Pipino, A., Lilly, S. J., Carollo, C. M. 2014, MNRAS, 441, 1444
\reference{}Pitts E., \& Tayler R. J. 1989, MNRAS, 240, 373
\reference{}Planck Collaboration 2018, ApJ, in press [arXiv:1807.06209]
\reference{}Popesso, P., Concas, A., Morselli, L., et al. 2019, MNRAS, 483, 3213
\reference{}Popping, G., Somerville, R.S., \& Galametz, M. 2017, MNRAS, 471, 3152
\reference{}Porter, L. A.; Somerville, R. S., Primack, J. R., et al. 2014, MNRAS, 445, 3092
\reference{}Recchi, S., \& Kroupa, P. 2015, MNRAS, 446, 4168
\reference{}Recchi, S., Spitoni, E., Matteucci, F., \& Lanfranchi, G.A. 2008, A\&A, 489, 555
\reference{}Remy-Ruyer, A., Madden, S.C., Galliano, F., et al. 2014, A\&A, 563, A31
\reference{}Renzini, A., \& Peng, Y.-J. 2015, ApJL, 801, L29
\reference{}Renzini, A. 2006, ARA\&A, 44, 141
\reference{}Ricarte, A., Tremmel, M., Natarajan, P., et al. 2019, MNRAS, submitted [arXiv:1904.10116]
\reference{}Rodighiero, G., Brusa, M, Daddi, E., et al. 2015, ApJL, 800, L10
\reference{}Rodighiero, G., Daddi, E., Baronchelli, I., et al. 2011, ApJL, 739, L40
\reference{}Rodriguez-Gomez, V., Pillepich, A., Sales, L. V., et al. 2016, MNRAS, 458, 2371
\reference{}Rodriguez-Gomez, V., Genel, S., Vogelsberger, M., et al. 2015, MNRAS, 449, 49
\reference{}Rodriguez-Puebla, A., Primack, J.R., Behroozi, P., \& Faber, S.M. 2016, MNRAS, 455, 2592
\reference{}Rodriguez-Puebla, A., Avila-Reese, V., Yang, X., et al. 2015, ApJ, 799, 130
\reference{}Romano, D., Karakas, A. I., Tosi, M., \& Matteucci, F. 2010, A\&A, 522, A32
\reference{}Saintonge, A., Catinella, B., Tacconi, L.J., et al. 2017, ApJS, 233, 22
\reference{}Sanders, R. L., Shapley, A. E., Kriek, M., et al. 2015, ApJ, 799, 138
\reference{}Santini, P., Maiolino, R., Magnelli, B., et al. 2014, A\&A, 562, A30
\reference{}Schaye, J., Crain, R.A., Bower, R.G., et al. 2015, MNRAS, 446, 521
\reference{}Schmidt, M. 1963, ApJ, 137, 758
\reference{}Schmidt, M. 1959, ApJ, 129, 243
\reference{}Schonrich R., \& Binney J. 2009, MNRAS, 396, 203
\reference{}Schreiber, C., Pannella, M., Leiton, R., et al. 2017, A\&A, 599, A134
\reference{}Scoville, N., Lee, N., Vanden Bout, P., et al. 2017, ApJ, 837, 150
\reference{}Scoville, N., Sheth, K., Aussel, H., et al. 2016, ApJ, 820, 83
\reference{}Scoville, N., Aussel, H., Sheth, K., et al. 2014, ApJ, 783, 84
\reference{}Shankar, F., Lapi, A., Salucci, P., de Zotti, G., \& Danese, L. 2006, ApJ, 643, 14
\reference{}Shi, J., Lapi, A., Mancuso, C., Wang, H., \& Danese, L. 2017, ApJ, 843, 105
\reference{}Shull, J. M., Harness, A., Trenti, M., \& Smith, B. D. 2012, ApJ, 747, 100
\reference{}Silk, J., \& Mamon, G.A. 2012, RAA, 12, 917
\reference{}Silk, J., \& Rees, M. J. 1998, A\&A, 331, L1
\reference{}Simpson, J. M., Smail, I., Swinbank, A. M., et al. 2015, ApJ, 807, 128
\reference{}Somerville, R.S., \& Dave, R. 2015, ARA\&A, 53, 51
\reference{}Somerville, R.S., Hopkins, P.F., \& Cox, T.J. 2008, MNRAS, 391, 481
\reference{}Sommariva, V., Mannucci, F., Cresci, G., et al. 2012, A\&A, 539, A136
\reference{}Speagle, J. S., Steinhardt,C. L., Capak, P. L., \& Silverman, J. 2014, ApJS, 214, 15
\reference{}Spilker, J. S., Marrone, D. P., Aravena, M., et al. 2016, ApJ, 826, 112
\reference{}Spitoni, E., Vincenzo, F., Matteucci, F. 2017, A\&A, 599, A6
\reference{}Spitoni, E., Romano, D., Matteucci, F., Ciotti, L. 2015, ApJ, 802, 129
\reference{}Spitoni, E., Matteucci, F., \& Marcon-Uchida, M.M. 2013, A\&A, 551, A123
\reference{}Spitoni, E., Recchi, S., \& Matteucci, F. 2008, A\&A, 484, 743
\reference{}Springel, V., Pakmor, R., Pillepich, A., et al. 2018, MNRAS, 475, 676
\reference{}Steidel, C.C., Rudie, G.C., Strom, A.L., et al. 2014, ApJ, 795, 165
\reference{}Steinhardt, C. L., Speagle, J. S., \& Capak, P. 2014, ApJL, 791, L25
\reference{}Stevens, A.R.H., Lagos, C.d.P., Obreschkow, D., \& Sinha, M. 2018, MNRAS, 481, 5543
\reference{}Stevens, A.R.H., \& Brown, T. 2017, MNRAS, 471, 447
\reference{}Stevens, A.R.H., Croton, D.J., \& Mutch, S.J. 2016, MNRAS, 461, 859
\reference{}Straatman, C. M. S., Labbe, I., \& Spitler, L. R. 2015, ApJL, 808, L29
\reference{}Sutherland, R. S., \& Dopita, M. A. 1993, ApJS, 88, 253
\reference{}Suzuki, T.L., Kodama, T., Onodera, M., et al. 2017, ApJ, 849, 39
\reference{}Tacchella, S., Bose, S., Conroy, C., Eisenstein, D.J., \& Johnson, B.D. 2018a, ApJ, 868, 92
\reference{}Tacchella, S., Carollo, C. M., Forster Schreiber, N. M., et al. 2018b, ApJ, 859, 56
\reference{}Tacconi, L. J., Genzel, R., Saintonge, A., et al. 2018, ApJ, 853, 179
\reference{}Tadaki, K.-I., Genzel, R., Kodama, T., et al. 2017a, ApJ, 834, 135
\reference{}Tadaki, K.-I., Kodama, T., Nelson, E. J., et al. 2017b, ApJL, 841, L25
\reference{}Talbot, R.J., Jr., \& Arnett, W. D. 1971, ApJ, 170, 409
\reference{}Talia, M., Pozzi, F., Vallini, L., et al. 2018, MNRAS, 476, 3956
\reference{}Thomas, D., Maraston, C., Schawinski, K., Sarzi, M., \& Silk, J. 2010, MNRAS, 404, 1775
\reference{}Thomas, D., Maraston, C., Bender, R., \& Mendes de Oliveira, C. 2005, ApJ, 621, 673
\reference{}Tinker, J., Kravtsov, A.V., Klypin, A., et al. 2008, ApJ, 688, 709
\reference{}Tinsley, B. M. 1974, ApJ, 192, 629
\reference{}Torrey, P., Vogelsberger, M., Marinacci, F., et al. 2019, MNRAS, 484, 5587
\reference{}Totani T., Morokuma T., Oda T., Doi M., \& Yasuda N. 2008, PASJ, 60, 1327
\reference{}Tremonti, C. A., Heckman, T. M., Kauffmann, G., et al. 2004, ApJ, 613, 898
\reference{}Valiante R., Schneider R., Bianchi S., Andersen A. C., 2009, MNRAS, 397, 1661
\reference{}Velander, M., van Uitert, E., Hoekstra, H., et al. 2014, MNRAS, 437, 2111
\reference{}Venemans, B.P., Decarli, R., Walter, F., et al. 2018, ApJ, 866, 159
\reference{}Vincenzo, F., Matteucci, F., \& Spitoni, E. 2017, MNRAS, 466, 2939
\reference{}Vincenzo, F., Matteucci, F., Belfiore, F., Maiolino, R. 2016, MNRAS, 455, 4183
\reference{}Vogelsberger, M., Genel, S., Springel, V., et al. 2014, MNRAS, 444, 1518
\reference{}Walcher, C.J., Yates, R.M., Minchev, I., et al. 2016, A\&A, 594, A61
\reference{}Wang, R., Carilli, C. L., Wagg, J., et al. 2008, ApJ, 687, 848
\reference{}Watson, W.A., Iliev, I.T., D'Aloisio, A., Knebe, A., Shapiro, P.R., \& Yepes, G. 2013, MNRAS, 433, 1230
\reference{}Weinberg, D.H., Andrews, B.H., \& Freudenburg, J. 2017, ApJ, 837, 183
\reference{}Whitaker, K. E., Franx, M., Leja, J., et al. 2014, ApJ, 795, 104
\reference{}White, S.D.M., \& Frenk, C.S. 1991, ApJ, 379, 52
\reference{}Willott, C. J., Bergeron, J., \& Omont, A. 2015, ApJ, 801, 123
\reference{}Wojtak, R., \& Mamon, G. A. 2013, MNRAS, 428, 2407
\reference{}Zahid, H. J., Kashino, D., Silverman, J. D., et al. 2014, ApJ, 792, 75
\reference{}Zhao, D. H., Mo, H. J., Jing, Y. P., \& Borner, G. 2003, MNRAS, 339, 12
\reference{}Zhukovska, S., Dobbs, C., Jenkins, E.B., Klessen, R.S. 2016, ApJ, 831, 147
\reference{}Zhukovska, S., Gail, H.-P., \& Trieloff, M. 2008, A\&A, 479, 453
\reference{}Zjupa, J., \& Springel, V. 2017, MNRAS, 466, 1625
\reference{}Zolotov, A., Dekel, A., Mandelker, N., \& Tweed, D. 2015, MNRAS, 450, 2327

\end{references}
\end{document}